# EVALUATION OF A COMPTON CAMERA SYSTEM FOR WHOLE-BODY IMAGING OF AC-225: A SIMULATION STUDY

Yuemeng Feng*, Arkadiusz Sitek, Shadi Abdar Esfahani, Hamid Sabet*, *Senior Member, IEEE*

***Abstract*—In this study, we introduce a Compton camera system design for whole-body imaging of Actinium-225 ($^{225}$Ac), one of the trending radionuclides for targeted alpha therapy (TAT). The system enables multi-energy gamma photon detection with higher efficiency compared to mechanically collimated SPECT. This design consists of two detectors made of Cadmium zinc telluride (CZT), providing a field of view (FOV) adequate for whole-body imaging, while achieving high sensitivity and clinically usable imaging resolution within a reasonable scanning time. This work focuses on the system design and evaluation using the Monte Carlo simulation toolkit Gate. The imaging performance is evaluated at two energy peaks (218 keV, 440 keV) by applying the energy windows of 211-225 keV and 430-450 keV, representing the major detectable gamma energies generated from $^{225}$Ac. We explore the possibility of using the Compton camera system for treatment response monitoring in TAT. The full decay chain of $^{225}$Ac is simulated. Results demonstrate an image resolution of 1.0 cm using a NEMA IQ phantom with 5.7 MBq of $^{225}$Ac simulated in a cold background. An image resolution of 1.3 cm can be achieved with a hot-to-background ratio of 30:1, and a resolution of 3.7 cm can be achieved with an activity ratio of 12:1. The best achievable sensitivity at a 10 cm distance from the detector is 0.5% at 218 keV and 0.45% at 440 keV as calculated using a pure gamma source in the simulation. The proposed system may serve as an alternative imaging tool for TAT scanning in clinical settings.**

## I. INTRODUCTION

We evaluate the imaging performance of a Compton scattering-based system for whole-body imaging of targeted alpha therapy (TAT) using the Monte Carlo simulation toolkit GATE [1]. The proposed system enables multi-energy gamma detection with higher efficiency compared to mechanically collimated SPECT. The concept of the Compton camera (CC) for clinical applications was first proposed in the 1970s ([2]). During photon detection, Compton scattering is utilized in Compton cameras, providing them with advantages in multi-energetic gamma imaging. The potential for distinguishing multi-energetic gamma rays, high sensitivity detection, and reduced acquisition time, along with the gantry simpler than conventional SPECT system, make CC a competitive non-invasive imaging methodology in both clinical ([3, 4]) and preclinical settings.

The feasibility of in vivo human imaging using a Compton system has been demonstrated in [5] where technetium-99m dimercaptosuccinic acid ($^{99m}$Tc-DMSA) and fluorine-18 fluorodeoxyglucose ($^{18}$F-FDG) are utilized, two of the most widely used tracers in SPECT and PET imaging. Evidence has shown that Compton cameras are capable of imaging these two tracers at clinical dose level. Benefiting from the collimator-free feature, a Compton scattering based prototype for image-guided surgery is proposed in [6]. Furthermore, a hybrid Compton-PET system is proposed in [7], with experimental outcomes demonstrating successful detection of multi gamma-ray accumulation in multiple organs for small animal imaging. While there is substantial evidence supporting the achievable Compton imaging resolution for multi-energetic gamma-ray imaging in small animals, there is currently no conclusive evidence regarding the potential resolution attainable with Compton camera for whole-body human imaging. Several research efforts are ongoing to enhance the imaging quality of Compton camera. These include new system geometry designs ([3, 8]), and accurate modeling in image reconstruction ([9, 10]). Unlike the linear projection employed in PET and SPECT systems, CC reconstruction is based on Compton conical projection, where angular uncertainties are crucial for evaluating the imaging resolution. Consequently, both the detector's intrinsic resolution and detector energy resolution should be considered in the system design, given that the Compton cone is defined by photon energies and the interaction locations.

Beyond its application for multi-tracer imaging in preclinical and clinical, CC imaging can serve as a tool for treatment monitoring during targeted radionuclide therapy (TRT). TRT has been explored for cancer treatment over the past decades and has gained increasing interest in clinical studies following the success of FDA-approved radionuclide: Radium-223 ($^{223}$Ra) dichloride ([11, 12]). Among ongoing studies, two types of radiation emitters are under investigation: alpha and beta particles. Alpha emitters are considered more effective than beta emitters due to their more efficient killing of malignant cells and less damage to normal tissue ([16]). With the growing

This work did not involve human subjects or animals in its research. The authors are with Department of Radiology, Massachusetts General Hospital, Harvard Medical School (e-mail: yfeng16, asitek, abdaresfahani.shadi, hsabet@mgh.harvard.edu).

development of TAT for cancer treatment, several alpha emitters such as Actinium-225 ($^{225}$Ac), Bismuth-213 ($^{213}$Bi), Astitine-211 ($^{211}$At), and Lead-212 ($^{212}$Pb), combined with antibodies or small molecules, are being investigated in clinical trials. During the decay of both alpha emitters and beta emitters, multiple energetic gammas are generated, enabling quantitative imaging of TRT. Developing a novel imaging system or adapting the current conventional imaging system for multi-energetic gamma quantification is essential for quality assurance and safety control in TRT. Mechanical collimation in SPECT offers advantages in multi-isotope imaging ([13]), however, its sensitivity and imaging resolution are lower than PET. The typical resolution for conventional SPECT applications for alpha imaging is 12 to 16 mm ([14]). PET provides better sensitivity than SPECT systems, but its detectable gamma rays are limited to two opposite gamma photons at 511 keV. The limitations of conventional SPECT system for multi-isotope imaging can be addressed by developing next generation of Cadmium zinc telluride (CZT) SPECT systems ([15-17]). For the widely used radioisotope $^{99m}$Tc, a novel CZT-SPECT (VERITON camera) has demonstrated a resolution better than 9.5 mm for cold rods recovery in a Deluxe Jaszczak phantom filled with 350 MBq of $^{99m}$Tc uniformly distributed in the background ([15]). In the context of targeted radionuclide therapy, quantitative imaging for beta emitters such as $^{177}$Lu using SPECT systems is validated ([18-21]). Novel CZT-SPECT scanners with multiple heads and a fixed gantry have shown advantages over the conventional SPECT for $^{177}$Lu imaging ([16, 17]). However, SPECT image quality for alpha-emitter imaging is lower than for beta emitters imaging ([22]). This is mainly because the injected dose of alpha emitters is much lower than that of beta emitters, resulting in fewer detectable gamma rays counts during decay. Additionally, the gamma rays emitted during the decay of alpha emitters such as $^{225}$Ac, often have higher energies than those produced during the decay of beta emitters, resulting in a greater likelihood of Compton scattering in the detector over photoelectric absorption.

The limitations of mechanically collimated SPECT in the context of TAT may be addressed by developing a Compton camera system. This study investigates the possibility of using a Compton camera for whole-body imaging, with $^{225}$Ac selected as the targeted radionuclide. The alpha agent $^{225}$Ac emits four alpha particles during the decay chain, making it more efficient when accurately targeted to the tumor compared to other alpha emitters. However, this also increases its toxicity if the radionuclide is over distributed to normal tissue. Over the past years, $^{225}$Ac has been used to treat various cancer types, and ongoing clinical trials continue to assess its dose tolerance and toxicity [23]. During the decay of $^{225}$Ac, alpha daughters such as $^{221}$Fr (11.6% emission probability) and $^{213}$Bi (26.1 % emission probability) are generated, which then decay emitting 218 keV and 440 keV photons detectable by gamma cameras ([24]). We have summarized several clinical cases in table 1. Depending on the patient's weight and the diseases, the administered dose might vary. The values presented in table 1 serve as a guide for our simulation's activity settings. The typical administered dose of $^{225}$Ac is 2 mCi (7.4 MBq) as summarized in [23].

TABLE I
$^{225}$Ac DOSE LEVEL IN THE TREATMENT OF CANCER

| Setting | Dose |
|---|---|
| **Neuroendocrine tumors** | 5.5 MBq-7 MBq [25, 26] |
| **Glioblastoma** | 10, 20, 30 MBq [27] |
| **Prostate cancer** | 100 kBq/kg-50 kBq/kg [28] |
| **Small cell lung cancer** | 120 kBq/kg [23] |
| **Acute myeloid leukemia** | 0.074-0.0555 MBq/kg [29] |

Currently, limited approaches have been proposed for conventional SPECT in TAT imaging. In [15], an imaging study of prostate cancer treatment with an injection of 5.27 MBq of $^{223}$Ra is demonstrated, where tumor uptake is visible after 30 minutes acquisition using the CZT-SPECT system. No data has yet been reported on the quantification of activity concentration in clinical cases with $^{225}$Ac. Another study has demonstrated the possibility of quantitative SPECT imaging for alpha agents ([22]) with three state-of-the-art SPECT/CT system (GE discovery NM/CT 670, GE Optima NM/CT 640, Siemens Symbia T6). The best sensitivity, when applying the energy windows of 216.8 keV $\pm$ 8%, 444.3keV $\pm$ 5% is around 50 cps/MBq for the Siemens Symbia T6 ([22]). The acquisition time is 30 seconds per projection acquired every 6 degrees over a 360-degree. Given that the total injection of $^{225}$Ac in clinical setting is as low as 5.5-30 MBq, the total detected gamma counts will be very limited in a realistic scenarios, compared to the most often used radionuclides for diagnostic imaging, such as $^{99m}$Tc. Due to the low gamma counts, the spheres smaller than 1.3 cm in the tested NEMA phantom are not visible with 35 kBq/mL of $^{225}$Ac, and only the largest (3.7 cm) and second largest (2.8 cm) spheres were used for quantitation in [22].

Enhancing sensitivity through the utilization of a Compton camera system could potentially address the issue of low detected counts. However, the image resolution may be lower compared to mechanically collimated SPECT when the detected photon energy is below 400 keV. In [30], by placing a GAGG-based Compton camera with 10×10 m$^2$ surface area at a distance of 30 cm from the patient injected with Ra$^{223}$, with targeting energy band at 324-351 keV, a 10-minute acquisition demonstrated the possibility of imaging the source distribution within the whole body of the patient. The results were compared to collimated SPECT, showing less advantage in image resolution, but successfully demonstrated a wider FOV and shorter acquisition time. The Compton camera imaging device MACACO III+ ([31]), which is based on lanthanum bromide (LaBr3) crystal and is currently under development, has demonstrated a resolution of 6 mm using experimental measurements with an $^{225}$Ac-filled Derenzo phantom. Improvements in angular resolution at 440 keV have been achieved by enhancing the spatial and energy resolution of the detector and increasing the dimensions of the detector plane.

The system design of Compton camera is one crucial aspect that needs to be investigated in order to achieve a good resolution. Although the primary focus of many studies is gamma ray imaging, as summarized in [32], rather than alpha monitoring, they have shown that image resolution improves with a larger distances between the scatterer and absorber, as well as with smaller pixel sizes. Moreover, a tenfold increase in sensitivity compared to Anger camera was demonstrated in [33] using a single-pixel scatterer with a larger surface absorber in a Compton camera for gamma imaging, and better image resolution compared to Anger camera with a limited FOV, can also be achieved, even with the same number of counts used in reconstruction. The MACACO III has recovered 4 mm diameter rods of a Derenzo phantom using $^{18}$F-FDG [34], and has demonstrated improved resolution when imaging a $^{131}$I-NaI phantom compared to conventional gamma camera. Compton imaging is particularly valuable in radiopharmaceutical development ([8]), where each experimental cycle often requires repeated small-animal scans. Over the past two decades, studies have successfully demonstrated the feasibility of small animal imaging using the Compton camera modality. However, the potential for human whole-body imaging remains uncertain. As targeted alpha therapy is a novel therapeutic approach under investigation, there exists no standard protocol for resolution measurement. Various studies have simulated different dosages. A GAGG-based Compton camera for small animal imaging was presented in [35], where the authors simulated hot rods with activities ranging from 3 to 8 MBq/cc in water. A CZT-based small animal imaging system with enhanced sensitivity was proposed in [8]. The authors simulated a smaller Derenzo phantom with activities of 4 μCi (0.148 MBq) and 1 μCi (0.037 MBq) for 15 minutes in air, the phantom activities were measured at 1.97 MBq/cc and 0.49 MBq/cc, respectively, achieving a best achievable resolution of 2 mm. However, despite the ongoing developments in medical imaging applications, fewer research groups are developing the Compton camera system specifically for whole-body alpha monitoring. Many groups have evaluated Compton camera systems for diagnostic imaging using Derenzo phantoms; however, these studies fall outside the scope of alpha imaging. It is important to note that alpha and gamma imaging serve fundamentally different purposes. Alpha imaging is used in the context of treatment monitoring, where the administered therapeutic doses result in relatively low-intensity gamma emissions, limiting the number of counts available for image reconstruction. In contrast, diagnostic gamma imaging typically involves higher-activity sources that yield more abundant emissions, enabling higher-quality image reconstruction. As a result, the simulation and acquisition conditions—particularly in terms of source activity—differ significantly between the two applications, leading to distinct image quality characteristics.

This paper explores the feasibility of utilizing a Compton camera for whole-body imaging in the context of $^{225}$Ac imaging. To evaluate the feasibility of using a Compton camera system for whole body imaging with $^{225}$Ac, we systematically assess various geometric parameters, including the thickness of the scatterer and absorber, the distance from the scatterer to the absorber, and the spatial resolution of the detectors. We employ reconstructed image resolution as key metric for imaging system performance. The objective is to determine the optimal geometric parameters for each detector that yield optimal performance. With these optimized geometry parameters, we evaluate the Compton camera system designed for human body imaging. The innovation of this study lies in the proposed imaging system, which features two detector heads: one fixed opposite the patient bed, and the other positioned parallel to the patient's side, movable in three dimensions. The design ensures both high sensitivity and clinically applicable imaging resolution. The orthogonal configuration enhances imaging resolution in the third dimension, while the increased opening area compared to two parallel planar detectors allows for a more comfortable positioning of the patient. We propose a concept for a device that facilitates treatment monitoring during targeted alpha therapy by imaging the gamma rays emitted during the daughter production of the treatment compound. Additionally, a single detector within the system can also be utilized for radio-pharmaceutical development by performing small animal scans.

## II.METHODS

This study evaluates the imaging performance of a Compton camera system for whole-body imaging using the 'CCMod' in Monte Carlo simulation toolkit Gate 9.2 ([36]). We first measure the achievable imaging resolution by simulating a Derenzo phantom and a point-like source, then simulate a NEMA IQ phantom to assess the contrast recovery ratio of the system. Section 2 (METHODS) presents the simulation setting, including the different geometry parameters simulated for detector design and the two-detector system. The reconstruction methods are also described in section 2. The reconstruction results of the single detector and two-detector system across various locations within the FOV is presented in section 3 (RESULTS). We discuss the performance of the imaging system and summarize both its limitations and advantages in section 4 (DISCUSSION). Lastly, we provide insights into the practicality and effectiveness of our current design for whole body imaging in TAT.

We simulated both single detector and double detector mode, with each mode comprising multiple subsets of simulations. We intend to explore the Compton imaging for targeted alpha therapy and specifically for the alpha emitter $^{225}$Ac. Besides $^{225}$Ac, we add additional simulations of Technetium-99m ($^{99m}$Tc) and Lutetium-177 ($^{177}$Lu) for evaluating the resolution of the Compton camera system, as $^{99m}$Tc is usually tested to measure the imaging performance of SPECT systems, and $^{177}$Lu is a popular beta emitter having been extensively investigated, and the quantifying imaging of $^{177}$Lu with state-of-art SPECT has been proven in many studies. Currently, the studies of conventional SPECT imaging with $^{225}$Ac is very limited, therefore we include $^{99m}$Tc and $^{177}$Lu in our tests to compare the performance of our system with others, by measuring the FWHM of the reconstructed point-like sources. We employed the list mode Maximum Likelihood Expectation Maximization (MLEM), reconstruction algorithm presented in [9] to evaluate the imaging resolution. In the reconstruction process, geometric sensitivity was determined using the methods proposed in [37]. For $^{225}$Ac imaging, point-like sources, Derenzo phantom, and a NEMA IQ phantom are simulated and reconstructed. The NEMA IQ phantom filled with $^{99m}$Tc was also simulated and reconstructed to compare image contrast recovery with other existing SPECT systems. The simulation settings of the source are presented in section II.A. The detector design is summarized in section II.B. The whole-body imaging system is detailed in section II.C. The image reconstruction methods are described in section II.D.

*A. Simulation setting: source*

The simulation in this study is divided into two parts: one for single-detector mode and one for the double-head system. The single-detector mode aims to identify the optimal parameters for the detector geometry. The parameters that lead to the best performance of the single detector will then be applied to the double-head system. The double-head system part focuses on evaluating the system's performance for whole-body imaging. With two main objectives—optimizing the detector parameters and evaluating the entire system—we conducted two sets of simulations, each with different sources, simulated in GATE.

*1. Single-Detector Mode*

The single detector mode aims to identify optimal geometry parameters for a single detector. The selected parameters are later applied to the double-head system. The simulated phantom and source are detailed in tables 2 and 3.

TABLE II SINGLE-DETECTOR MODE: 2D DERENZO PHANTOM

| Parameter | Description |
|---|---|
| Objective | Optimize geometry parameters based on image resolution |
| Phantom length | 1 cm |
| Activity Ratio | 0.84 MBq/cc (hot rods), no background activity |
| Radionuclide | $^{225}$Ac |
| Simulated Peaks | 218 keV ($^{221}$Fr), 440 keV ($^{213}$Bi) |
| Total Activity | 15.1217 MBq |
| Acquisition Time | 16 minutes (in air) |
| Branching Ratios (%) | 11.4% (218 keV), 25.9% (440 keV) |
| Gamma Activity (MBq/cc) | 0.095 (218 keV), 0.218 (440 keV) |
| Half-Lives Simulated | 4.8 min ($^{221}$Fr), 45.6 min ($^{213}$Bi) |
| Source Location | Fixed at 10 cm to the scatterer |

TABLE III SINGLE-DETECTOR MODE: POINT-LIKE SOURCE

| Parameter | Description |
|---|---|
| Objective | Validate parameters optimized using Derenzo phantom |
| Radionuclides | $^{99m}$Tc, $^{177}$Lu, $^{225}$Ac (full decay chains) |
| Acquisition Time | 1 hour |
| Total Primaries | $10^6$ (in water) |
| Source Positions (cm) | (0,0,10), (0,0,15), (0,0,20), (0,10,10), (10,10,10) |

*2. Double-Head System*

The double head mode evaluates the imaging performance of the full system with both detectors fixed. The simulated phantom and source are detailed in table 4, 5 and 6.

TABLE IV DOUBLE-HEAD SYSTEM: POINT-LIKE SOURCE

| Parameter | Description |
|---|---|
| Objective | Assess resolution improvement from second detector |
| Radionuclides | $^{99m}$Tc, $^{177}$Lu, $^{225}$Ac (full decay chains) |
| Acquisition Time | 1 hour |
| Total Primaries | $10^6$ (in water) |
| Source Distance | 10 cm from each detector |

TABLE V DOUBLE-HEAD SYSTEM: 3D DERENZO PHANTOM

| Parameter | Description |
|---|---|
| Objective | Assess resolution within various locations of FOV |
| Phantom Length | 8 cm |
| Activity Ratio | 0.1 MBq/cc (hot rods), no background activity |
| Radionuclide | $^{225}$Ac |
| Simulated Peaks | 218 keV ($^{221}$Fr), 440 keV ($^{213}$Bi) |
| Total Activity | 14.4016 MBq |
| Source Location | Moved within system FOV |

TABLE VI DOUBLE-HEAD SYSTEM: NEMA IQ PHANTOM

| Parameter | Description |
|---|---|
| Objective | Evaluate detectability and recovery coefficients (RCs) |
| Hot Sphere Diameters (cm) | 3.7, 2.8, 2.2, 1.7, 1.3, 1.0 |
| Cold Region | 18 cm rod, 2.5 cm diameter |
| $^{225}$Ac Simulation - Total Activity | 5.7 MBq |
| $^{225}$Ac Simulation - Acquisition Time | 15 minutes |
| $^{225}$Ac Simulation - Activity Ratios | No background (30 kBq/ml), 30:1 (3.8 kBq/ml), 12:1 (1.6 kBq/ml) |
| $^{99m}$Tc Simulation - Total Activity | 320 MBq |
| $^{99m}$Tc Simulation - Acquisition Time | 10 minutes |
| $^{99m}$Tc Simulation - Activity Ratio | 8:1 (160 kBq/ml) |

For the simulated NEMA phantom filled with $^{99m}$Tc in water (table 5), the target to background concentration ratio for the hot spheres is simulated as 8:1, with the hot sphere concentration set to 160 kBq/ml. For the simulated NEMA phantom filled with $^{225}$Ac in water (table 5), three different target to background activity ratios are tested: no background, 30:1 and 12:1. The values 30:1 and 12:1 were chosen based on a study of the ex vivo time-dependent biodistribution of $^{225}$Ac in a mouse tumor model [38]. As there is not yet sufficient data on the biodistribution of $^{225}$Ac in patients, we used studies on small animal model as a guide for simulating the activity ratio. Studies on the biodistribution of $^{225}$Ac using mouse model have demonstrated its accumulation in targeted tumor and clearance in other organs, such as bone, blood, and liver [38-40]. The accumulation in the kidney is high within the first 24 hours post-injection, resulting in a tumor to kidney ratio as high as 1. For more details on the evaluation of the $^{225}$Ac-labeled antibody that accumulated in different organs within the mouse model, please refer to the data shown in [38]. Since $^{225}$Ac is used for therapy rather than diagnostic imaging, and has half-life of approximately 9 days, the observed accumulation in kidney underscores the need for careful evaluation of toxicity in clinical setting. The development of pharmaceuticals is beyond the scope of this study, and we conservatively chose the 12:1 hot region to background ratio as the lowest ratio in the simulation to evaluate the imaging performance for $^{225}$Ac. The total simulated dose of $^{225}$Ac in the NEMA phantom is 5.7 MBq, with an acquisition time of 15 minutes. The hot spheres concentration in the cold background is 30 kBq/ml. For the target to background ratio 30:1, the hot concentration is 3.8 kBq/ml, and for the ratio of 12:1 it is 1.6 kBq/ml. This study is based on Monte Carlo simulation; thus, the system calibration is not evaluated in this work.

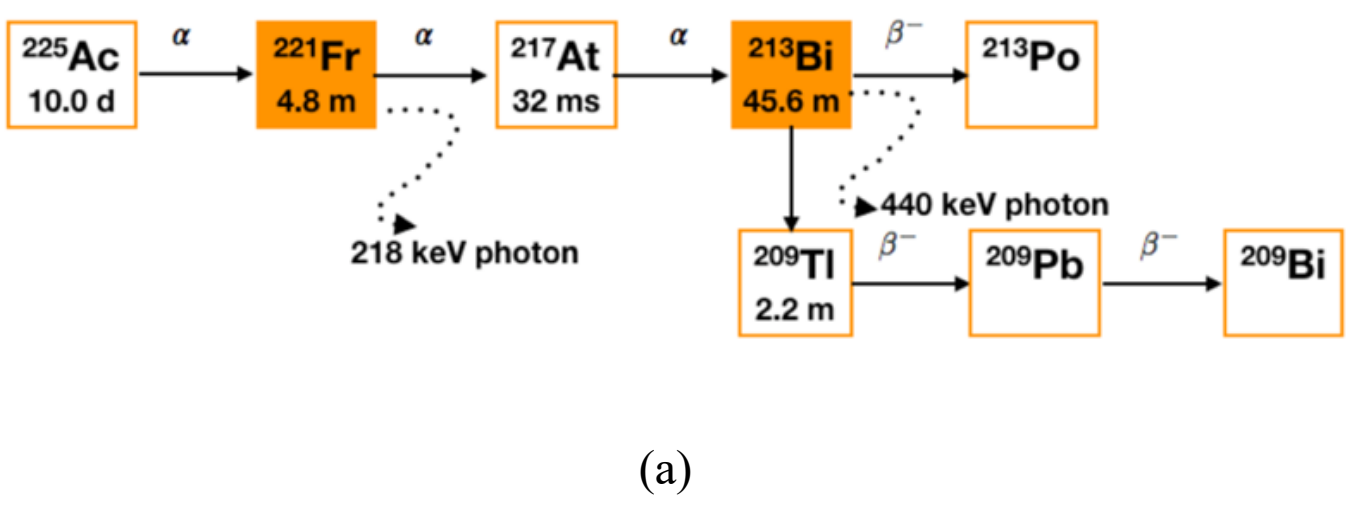

(a)

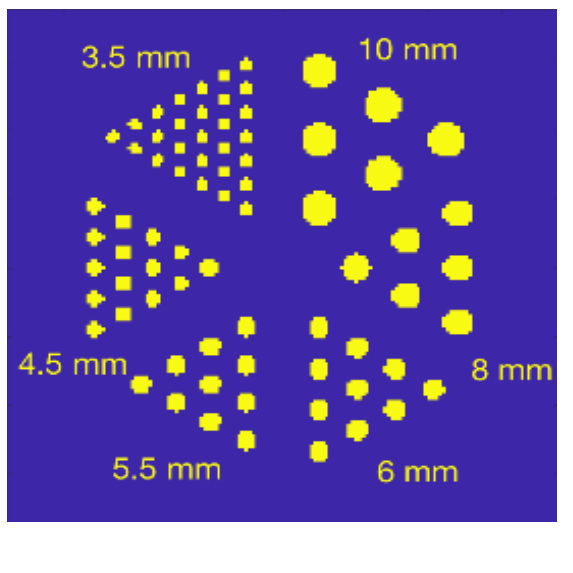

(b)

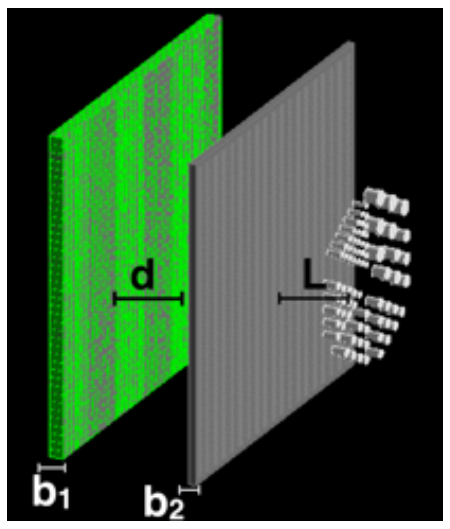

(c)

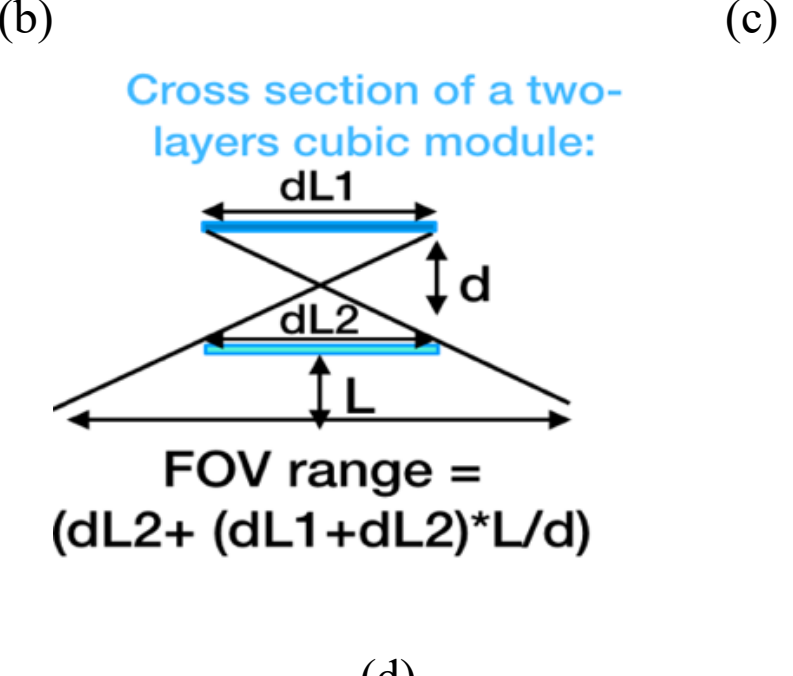

(d)

**Fig. 1.** (a) Decay chain of $^{225}$Ac, (b) central slice of the simulated Derenzo phantom, (c) simulation geometry of a single detector for parameters optimization with a 1 cm thick phantom, the source-to-detector distance is noted as L, the distance between the scattering layer and the absorption layer is noted as d, the thickness of scattering layer is noted as $b_2$, and the thickness of absorber is noted as $b_1$, (d) FOV calculation, $dL_1$ and $dL_2$ correspond to the size of scatterer and absorber borders.

### *B. Single Detector design*

The goal of this section is to determine the optimal geometry parameters for the detector shown in figure 1 (c). The single detector of the camera consists of one absorber and one scattering layer. Both layers have the same surface area on the detector but vary in thickness. A fixed 20 cm×20 cm detector surface is utilized. An illustration of the FOV calculation is shown in figure 1 (d). Denoting the distance from the scatterer to absorber as *d*, the size of the edge of the FOV at a distance *L* from the detector is calculated as $40 \times L/d + 20$ cm, with both the scatterer and absorber surface measuring 20 cm × 20 cm. To achieve an FOV size of 40 cm, equivalent to the average human shoulder width, at a distance of 10 cm from the source to the detector (half of the torso size in the sagittal plane), the distance *d* from the scatterer to the absorber should be less than 20 cm.
One critical factor influencing Compton imaging quality is the detector energy resolution, which contributes to angular uncertainties of the Compton cone. Various models have been proposed for calculating angular uncertainties based on energy measurements [41]. In this work, the simulated camera consists of two layers, one for scattering and one for photoelectric absorption. During the detection of photons, multi-scattering can occur. In the simulation, multi-scattering events and back scattering events are simulated, but multiple interactions events within the same detector volume are excluded. Only coincidences where the total deposited energies fall within an energy window ($E_0 \pm$ FWHM) and having interactions on two different detector layers are used in the reconstruction, with $E_0$ representing the initial energies of gamma rays.

An ideal detector material for a Compton camera should exhibit significant Compton scattering (CS) in the scattering layers and efficient photoelectric absorption (PEA) in the absorber, while also maintaining good energy resolution. The likelihood of Compton scattering interaction is proportional to the detector material density; however, materials should also allow scattered gammas to escape towards the absorber. In this study, CZT was selected as the scatterer due to its superior energy resolution and large count rate in room temperature with existing fabrication processes. The best achievable energy resolution at 122 keV ranges from 2.84% to 3.27% for CZT [42], considering manufacturing constraints, we chose a simulated energy resolution of 3.1% full width at half maximum (FWHM) at 218 keV. Both measurement-induced energy uncertainties and Doppler broadening are simulated. The physics list used in the simulation is 'emstandard' in Geant4 ([43]). The low timing resolution of CZT affects event selection in the Compton camera, however, the low activity of gamma rays generated from the decay chain of $^{225}$Ac reduces its impacts. A timing resolution of 10 ns is previously reported for a 1 cm thickness CZT crystal ([44]). In this study, we have not implemented timing correction. Coincidences are selected within a timing window of 3 ns in the simulation. We used 'CCMod' in Gate to simulate the system [36], the same module for processing coincidences in PET systems is applied, including dead time and memory buffer effects. Coincidences with total energy deposition within the ranges 211-225 keV and 430-450 keV are chosen in the reconstruction. In addition to the choice of detector material, energy resolution and detector surface, the following parameters were optimized in the system design to achieve optimal resolution and relatively high sensitivity.

1. scatterer thickness $b_2$
2. distance between scatterer and absorber $d$
3. scatterer pixel dimension $p_1$ and absorber pixel dimension $p_2$

Three scatterer thickness $b_2$ of 2 mm, 5 mm, and 10 mm are simulated, along with three scatterer-to-absorber distances $d$ of 10 cm, 15 cm, and 20 cm. Additionally, three detector resolutions are tested: 0.2 mm, 0.5 mm, 1.0 mm for the scatterer, and 0.5 mm, 1.0 mm, 1.5 mm for the absorber. The depth of interaction (DOI) information is not evaluated in this study. The detected position in depth with measurement uncertainties equal to the pixel size is used for locating the events.

The parameters mentioned above are investigated individually. The objective is to define the parameters that can achieve both good sensitivity and resolution. Reconstructed image resolution and detected counts are considered as metrics for assessing system performance. Reconstruction is conducted using the list-mode MLEM, with system matrix modeling following the approach proposed in [9]. All reconstructed central slices of the Derenzo phantom are obtained at the 20$^{th}$ iteration of the reconstruction with a voxel size set to 1 mm$^3$. We chose the results after 20 iterations because excessive iterations degrade image quality under the simulated limited-count conditions. Supporting evidence is provided in the Results section.

After selecting the optimized detector geometric parameters by using the Derenzo phantom, a point-like source was simulated separately at different locations and filled with different radionuclides ($^{225}$Ac, $^{177}$Lu, $^{99m}$Tc) using the optimized detector parameters. We report the FWHM of the reconstructed point source containing $^{99m}$Tc, $^{177}$Lu, $^{225}$Ac, each simulated separately in water. The following energy windows are applied for event selection in the reconstruction: for $^{99m}$Tc: 135.5-147.5 keV; for $^{177}$Lu: 204-212 keV; and for $^{225}$Ac: 211-225 keV and 430-450 keV.

The Compton scattering angle for a coincidence involving only two interactions, the first being Compton scattering and the second being photoelectric absorption, can be expressed by the following equation:

$$cos(\beta) = 1 - \frac{m_e c^2 E_1}{E_2(E_1+E_2)} \quad (1)$$

where $E_1$, $E_2$ are the energies deposited in the detector during the first and second interactions, respectively.

The energy detection model used to estimate the resolution of $\beta$ assumes that the uncertainties of energy detected by the scatterer and absorber both impact the Compton cone. The conical uncertainty $\rho$ of the cone intersection on the plane parallel to the detector at a distance $L$ can be approximately calculated by $\rho = \Delta\beta L$. The angular uncertainty model proposed in [41] is applied to calculate $\Delta\beta$ (eq. 2). The simulated energy resolution $R$ at energy $E$ follows inverse square law: $R = R_0 \frac{\sqrt{E_0}}{\sqrt{E}}$, where $R_0$ corresponds to the energy resolution at energy $E_0$. This law is applied in Gate simulation by default.

$$\Delta\beta = \frac{m_e c^2}{(E_1+E_2)^2 sin\beta} \sqrt{(\Delta E_1)^2 + \left(E_1(2E_2+E_1)\Delta E_2/{E_2}^2\right)^2} \quad (2)$$

Here, $E_1, E_2$ correspond to the measured energies in scattering and photoelectric absorption, and $\Delta E_1$, $\Delta E_2$ represent the uncertainties of these measurements. Doppler broadening is not counted in this equation but is accounted in the simulations of this work and is modeled as a Gaussian kernel in the reconstruction. In human whole-body imaging, the FOV is larger compared to small animal imaging, leading to significant variations in conical shell uncertainties within the FOV, which notably impact the reconstructed image resolution.

The absorber is designed to stop the scattered photons, and the thickness (unit in centimeters) required to stop 85% of photons at energy $E$ is calculated by

$$b = -10 * ln(1 - 0.85)/(G * k) \quad (3)$$

where $k$ is the material density, which is 5.83 $g\ cc^{-1}$ for CZT ([45]), and $G$ is the photoelectric absorption coefficient, for an incident photon energy $E$ of approximately 218 keV, $G$ is approximately 0.246 $cc\ g^{-1}$, a thickness of absorber larger than 13.2 mm can stop 85% photons. With an absorber thickness equal to 10 mm, as simulated in this study, approximately 76% of photons at 218 keV can be stopped and fully absorbed.

The scatterer thickness should be optimized to maximize the number of photons undergoing only Compton scattering without subsequent photoelectric absorption. When Compton scattering occurs first on the absorber followed by photoelectric absorption in the scattering layer, a backscattering event could be recorded and generated false Compton cone. During photon detection, without involving time-of-flight (TOF) techniques, both scattering and backscattering can occur, resulting in a higher number of detected photons than forward scattering counts. The possibility of Compton scattering and photoelectric absorption depends on the photon energy and material. We chose to determine the optimized scatterer thickness through Monte Carlo simulation instead of implementing approximate calculation. A point-like source is simulated separately at 218 keV and 440 keV in air and is emitting from a distance of 10 cm from the scatterer, with varying thickness of scatterer. The sensitivity is calculated as the ratio of detected coincidences to the total emitted counts. The distance between the scatterer and absorber is fixed at 10 cm, and the absorber thickness is fixed at 10 mm. Only the events having interactions on different detector volumes are considered in the calculation of sensitivity, no energy threshold was applied to this estimation of sensitivity, multi-scattering events on different layers are kept. Events having only one interaction, having multi-interaction on one detector volume are all rejected in the coincidence calculation, and not considered for calculating the sensitivity.

### *C. Double detector: Compton camera whole-body system, fixed mode*

Once the optimized geometry parameters are determined, the performance of the Compton camera system, comprising the optimized detector, is evaluated through simulation. The simulated system consists of 2 detector modules. Each module has 10×10 CZT blocks, where each block is composed of one layer for photoelectric absorption and one layer for Compton scattering. The absorber layer has 40×40 pixels, each measuring 0.5 mm × 0.5 mm with a thickness of 1 cm. The scattering layer has 100×100 pixels, each measuring 0.2 mm×0.2 mm with a thickness of 5 mm. The detailed simulation geometry of the system with a patient bed, is illustrated in figure 2 (a). Regardless of whether the simulated source was a Derenzo or a NEMA IQ phantom, it was centered at the red point shown in Figure 2(a) in simulation for evaluating the double-detector system. Denoting the distance from the source to the first and second detectors as $L_1$ and $L_2$, and using the FOV calculation illustrated in figure 1 (d), the length of the FOV edge can be computed as $dL_1 + \frac{(dL_1+dL_2)L_1}{d}$, where d is the absorber to scatterer distance set to 10 cm, and $dL_1$, $dL_2$ are 20 cm by the design. By applying a factor of $\sqrt{2}$ to account for the longest diagonal, and varying $L_1$ from 10 cm to 20 cm, the FOV size in the coronal plane ranges from 84 cm to 141 cm. This configuration provides adequate coverage for human torso in the coronal plane shown in figure 2(a), and with the two detectors movable along the patient bed, the whole system offers adequate coverage of the whole-body scanning.

We separately simulated point-like sources containing $^{225}$Ac, $^{177}$Lu and $^{99m}$Tc. Each source was placed 10 cm from both detectors. We compared the reconstruction of the point sources simulated with double detectors to those with single detector, to demonstrate the resolution improvements provided by the second detector. To evaluate the imaging performance at different locations within the FOV, we fixed the two detectors and measured the reconstructed image resolution by varying the distance from the detectors to a modified 8 cm long Derenzo phantom. We then simulated a NEMA IQ phantom. The simulation geometry of the Derenzo phantom is shown in figure 2 (b), the simulated NEMA IQ phantom is shown in figure 2 (c) and figure 2 (d). The closest side of the phantom to detector 2 is 5 cm. At this distance, detector 2 has a field of view (FOV) of 40 cm in the direction parallel to the phantom's height. Given that the phantom has a height of 18 cm and a width of 23 cm, it is entirely covered by detector 2, and its volume accounts for 19.7% of detector 2's coverage. For detector 1, the phantom is only partially covered. The closest side of the phantom is 1 cm from detector 1, which has a 24 cm FOV at this distance. As a result, the phantom is 99% covered by detector 1.

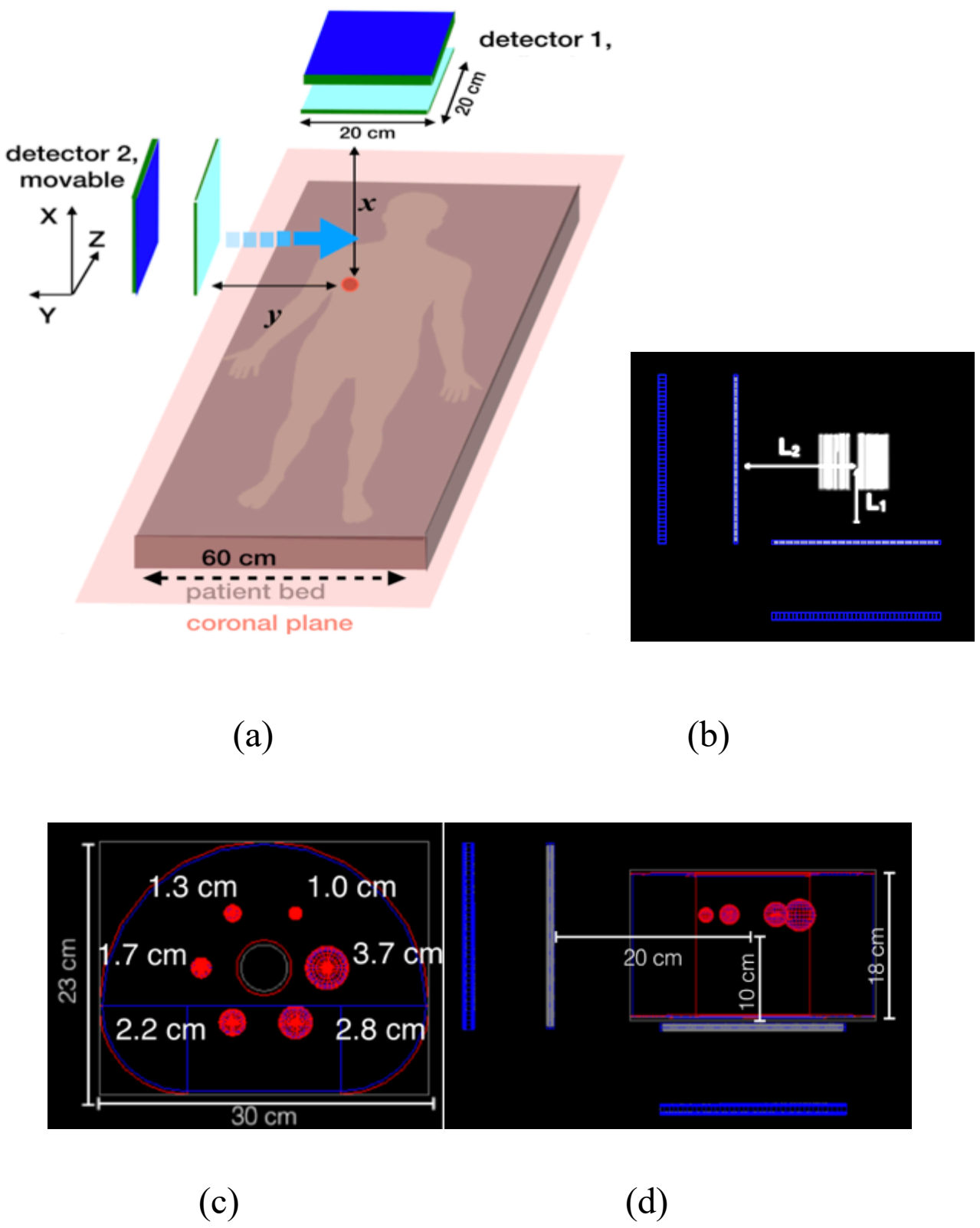

**Fig. 2.** (a) The 2 detectors together cover a whole-body imaging FOV. The red point corresponds to the center of the source, and the blue arrow indicates the possible direction of movement of detector 2, (b) simulation geometry of double detector with an 8 cm-long Derenzo phantom located at the center of FOV, with source-to-detector distances denoted by $L_1$ and $L_2$, (c) the cross section of the simulated NEMA IQ phantom, (d) simulation geometry with the NEMA

phantom placed 10 cm from the first detector, 20 cm from the second detector.

### *D. Image reconstruction methods*

All the reconstructed images shown in this manuscript are reconstructed using list mode MLEM proposed in [9], which estimates the photons emitted from the voxel j, $\lambda_j$, by iteratively calculating the sequence:

$$\lambda_j^{\,l+1} = \frac{\lambda_j^{\,l}}{s_j} \sum_{i=1} \frac{t_{ij}}{\sum_{k=1} t_{ik} \lambda_k^l}, \quad (4)$$

where *l* presents the current iteration number. The system matrix element $t_{ij}$, representing the probability that a photon emitted in voxel $v_j$ is detected as event i, and is calculated by the following equation:

$$t_{ij} = \int_{M \in v_j} K(\beta_G | E_0) \frac{\cos(\theta_{\overrightarrow{V_1 M}})}{|\overrightarrow{V_1 M}|^2} h(\delta | \beta_E) dv, \quad (5)$$

where the first term K corresponds to the Klein-Nishina differential cross section of Compton scattering with angle $\beta_G$ at initial energy $E_0$. The second term in equation 5 is the solid angle, M corresponds to the center of one voxel of the reconstructed volume, $V_1$ corresponds to the first interaction position. $\theta_{\overrightarrow{V_1 M}}$ is the angle of the first interaction on the scatterer. The third term represents the uncertainties of the aperture angle of the cone of response, which is a mixed Gaussian kernel estimating the uncertainties of Compton angle, in which both energy resolution and Doppler broadening is counted, $\delta$ *is calculated as* $\delta = \beta_G - \beta_E$, $\beta_E$ is the energetic scattering angle, $\beta_G$ is the geometric angle between the vectors $\overrightarrow{V_1 M}$ and $\overrightarrow{V_2 V_1}$, $V_2$ is the second interaction position. The methods for calculating the mixed Gaussian kernel was proposed in [37]. Figure 3 illustrates the parameters in equation 5.

The $S_j$ in equation 4 represents the probability that a photon emitted from voxel j is detected by the scanner, which is one element of the sensitivity matrix, and is calculated by considering the solid angle of the scatterer and the linear attenuation coefficient for Compton scattering using the following equation:

$$S_j \propto \iint_0^{dL} \frac{\cos\theta_j}{(x-x_j)^2 + (y-y_j)^2 + D_j^{\,2}} \left(1 - e^{-\mu_c b_2 / \cos\theta_j}\right) dydx, \quad (6)$$

where the first factor accounts for the solid angle of the detector pixel (x, y) seen from the center of the voxel, $D_j$ is the distance from the center of the voxel to the scatterer, dL is the length of the scatterer; the second term accounts for the probability of interaction in the scatterer, $b_2$ is the thickness of the scatterer, $\mu_c$ is the linear attenuation coefficients for Compton, $\theta_j$ is the angle between the voxel and detector pixel. Details on modeling of system matrix and sensitivity matrix and the validation of calculation against Monte Carlo simulation are given in previous studies [9] and [37].

The image reconstruction volume is 15×15×1 cm for the 1 cm-long Derenzo phantom, 15×15×10 cm for the 8 cm-long Derenzo phantom, 23×30×18 cm for the NEMA IQ phantom. The voxel size is 1 $mm^3$ for the reconstruction of Derenzo phantoms and point-like sources, and 2.5 $mm^3$ for the NEMA phantom. The images of Derenzo phantoms are shown at the 20$^{th}$ iterations. The images of NEMA IQ phantom are shown at the 50$^{th}$ iteration. The FWHM of the reconstructed point is measured at the 200$^{th}$ iteration.

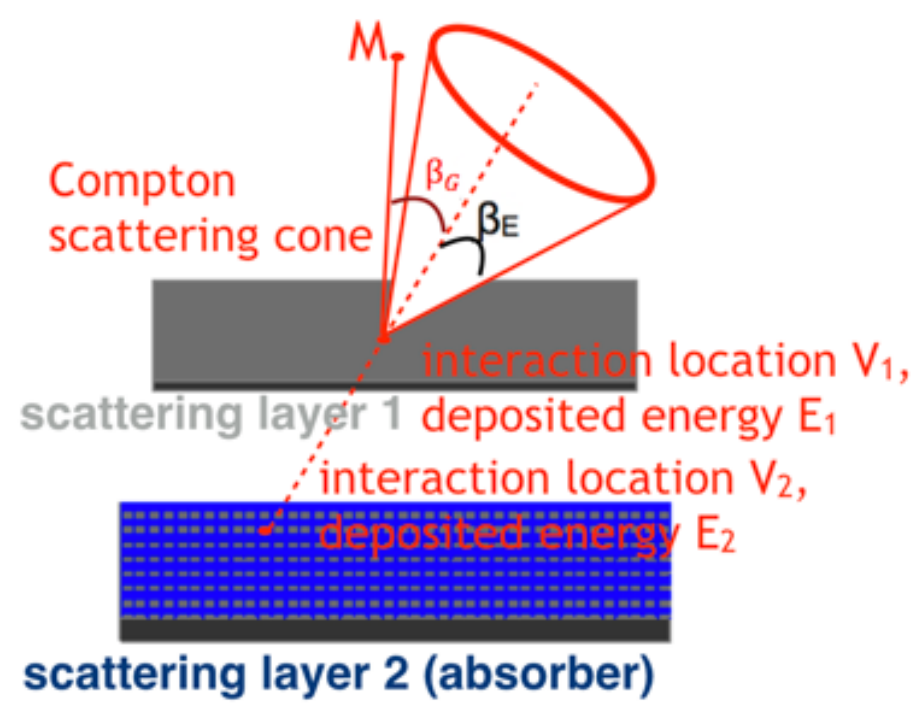

**Fig. 3.** Calculation of the Compton cone.

To evaluate the contrast recovery and activity recovery of the spheres in NEMA IQ phantom, we use the recovery coefficient (RC) and contrast recovery coefficient (CRC) calculated using the following equations:

$$RC_i = H_i / H_{t,i} \quad (7)$$

$$CRC_i = (H_i - B)/(H_{t,i} - B_t) \quad (8)$$

where $H_i$ is the measured concentration of hot region *i*, B is the measured concentration of the warm background, $H_{t,i}$ is the simulated hot concentration, $B_t$ is the simulated background concentration.

In the reconstruction process, only events that involve interactions across different layers of the detector within a timing window, with total deposited energies within an energy window, are classified as Compton coincidences. Multi-scattering events are filtered, photons that undergo multiple interactions within the same layer of the detector are excluded.

## III. Results

### *A. Sensitivity and resolution of single detector*

The simulated sensitivity at a distance of 10 cm to the detector with varying scatterer thicknesses is presented in figure 4. The proportion of multi-scattering is summarized in table 7 and 8, with simulated 5 mm thick scatterer.

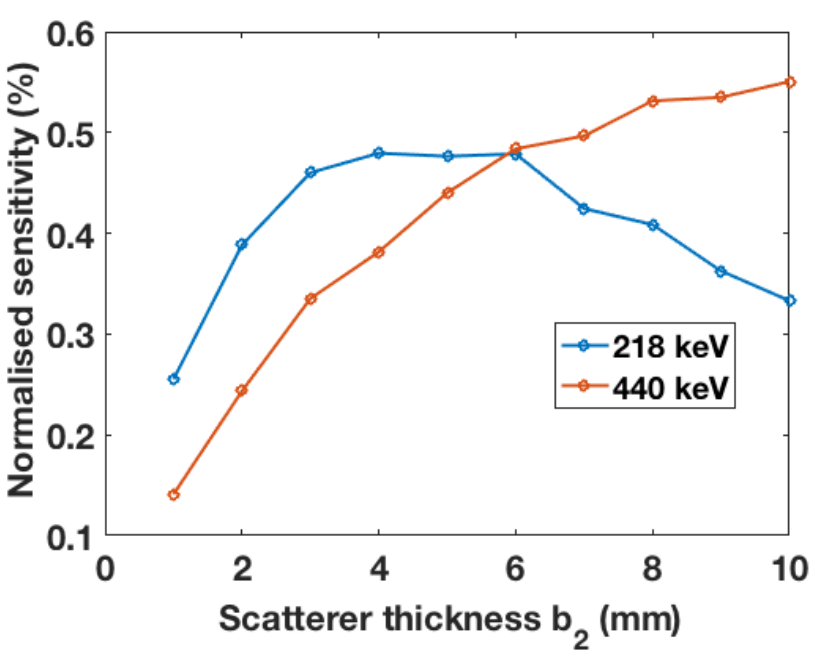

**Fig. 4.** Simulated sensitivity as a function of scatterer thickness at a distance of 10 cm from the detector.

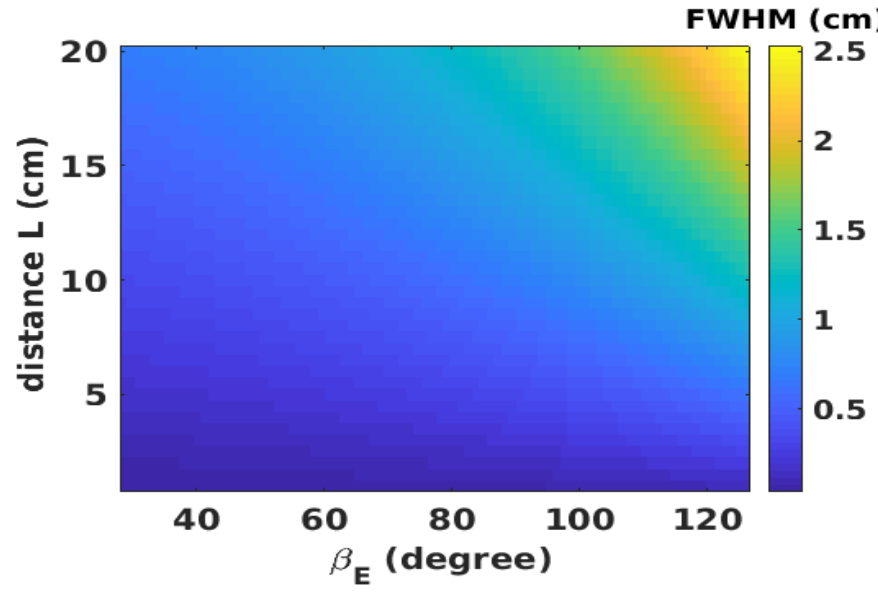

**Fig. 5.** FWHM at 218 keV of the uncertainties in the parallel cross-section of the Compton scattering cone, assuming an energy resolution of 3.1%, figure is plotted as a function of scattering angle $\beta_E$ and distance to the detector plane L, $\beta_E$ is determined based on the energies.

We plot the achievable FWHM of the conical uncertainties ρ as a function of angular uncertainties and distance to the detector, with the angular uncertainties expressed in equation 2. Figure 5 illustrates the FWHM of the uncertainties of the cross-section of Compton scattering cone as a function of scattering angle and distance to the detector plane, assuming an energy resolution of 3.1% at 218 keV for CZT. The energies $E_1$ and $E_2$ are supposed to sum up to the photon initial energy (218 keV) in figure 5, and the uncertainties of measurements $\Delta E_1, \Delta E_2$ are simulated using the inverse square law. The color-bar represents the FWHM in centimeters. The FWHM of the conical uncertainties ranges from 0.68 cm to 2.53 cm at a distance of 20 cm from the detector, with an average FWHM of 1.29 cm, calculated as the root mean square of the FWHMs. The actual reconstructed image resolution is better than 1.0 cm at a distance of 10 cm if the reconstruction has sufficient counts. With optimized parameters of the detector geometry, the best achievable resolution for the reconstructed images of the Derenzo phantom is 4.5 mm.

The reconstruction results of the Derenzo phantom at the 20th iteration are shown in figure 6 and figure 8 with varying scatterer-to-absorber distance d and varying scatterer thickness $b_2$, in both cases, at least 5.5 mm rods can be resolved. We chose 20 iterations because too many iterations will introduce high frequency noise without the recovery of the hot rods, the reconstruction is stopped at the 20th iteration to achieve smoother results. Please refer to figure 7 for the variation of CRC as a function of list mode MLEM iterations corresponding to the results shown in figure 6. The coincidence counts applied in the reconstruction for each geometry is summarized in table 9 and table 10. The detector pixel resolution is set to small value (FWHM of 0.2 mm for the scatterer and 0.5 mm for the absorber) to accurately estimate the detection location. Denoting the scatterer pixel resolution as $p_1$, absorber pixel resolution as $p_2$, the smallest diameter of rods resolved at the 20th iteration with varying detector pixel sizes is summarized in table 11 and table 12. Reconstructed images with various detector pixel sizes are shown in figure 11 and figure 12. The scatterer-to-absorber distance is fixed at 10 cm, the scatterer thickness is fixed at 5 mm. The line profiles of reconstructed rods are shown in figure 9, 10, 13, 14. The line profiles are normalized by the maximum of each reconstructed image. According to the results with varying geometry parameters, the optimized geometry parameter of the detector is selected as follow: scatterer thickness of 5 mm, absorber thickness of 10 mm, distance from scatterer to absorber 10 cm, the scatterer resolution is 0.2 mm and absorber resolution is 0.5 mm.

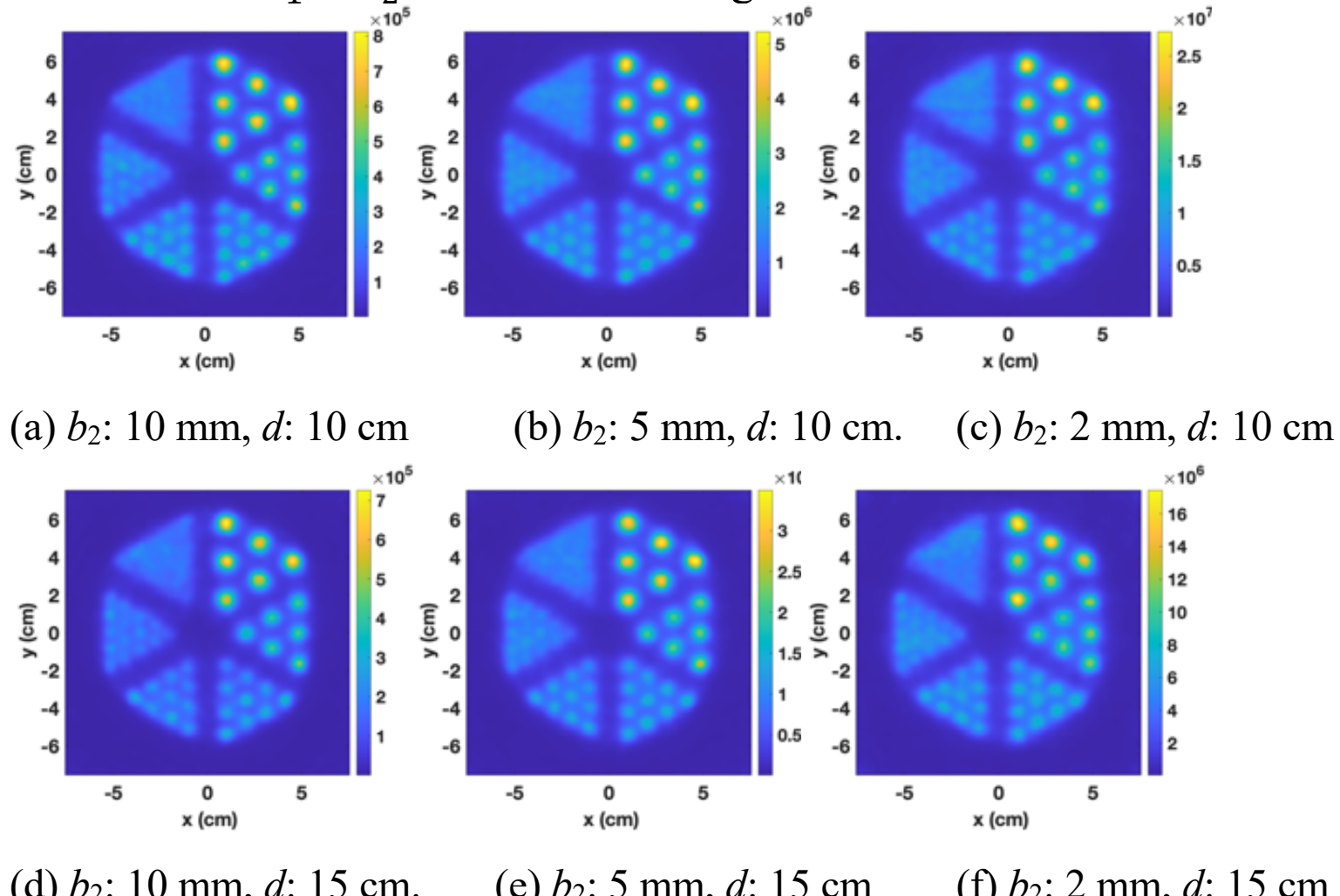

(a) $b_2$: 10 mm, $d$: 10 cm (b) $b_2$: 5 mm, $d$: 10 cm. (c) $b_2$: 2 mm, $d$: 10 cm

(d) $b_2$: 10 mm, $d$: 15 cm. (e) $b_2$: 5 mm, $d$: 15 cm (f) $b_2$: 2 mm, $d$: 15 cm

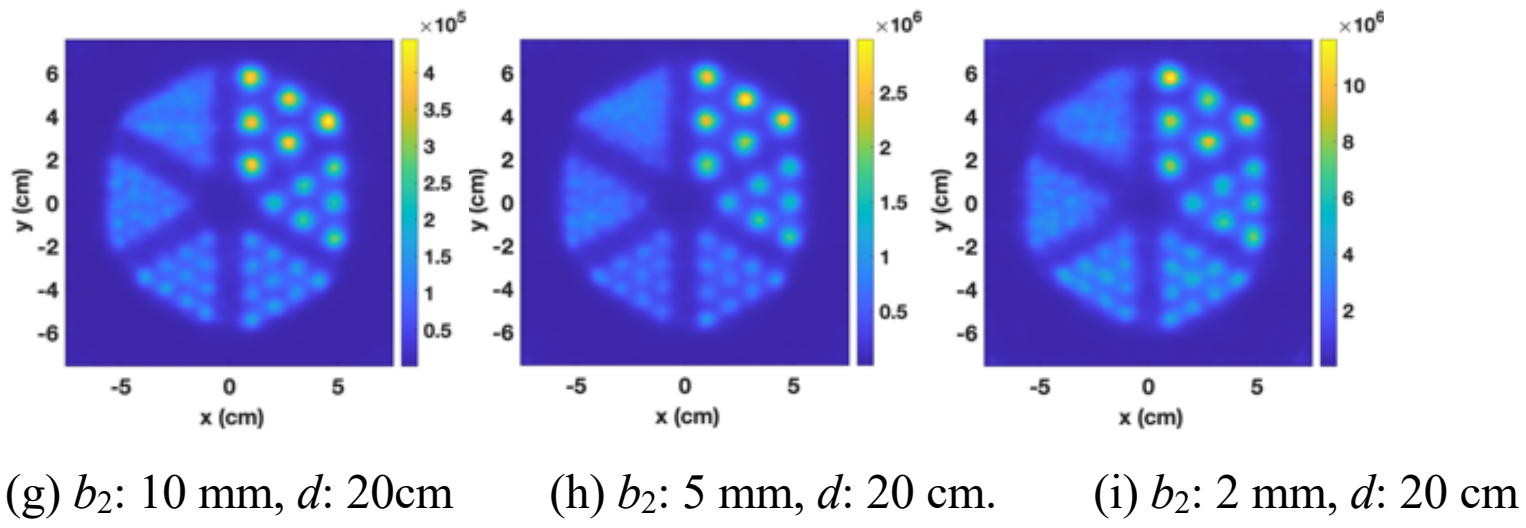

(g) $b_2$: 10 mm, *d*: 20cm (h) $b_2$: 5 mm, *d*: 20 cm. (i) $b_2$: 2 mm, *d*: 20 cm

**Fig. 6.** Reconstruction results at 218 keV with the source placed at 10 cm parallel to the detector plane. Simulations were conducted using three scatterer thicknesses $b_2$ (2 mm, 5 mm, 10 mm) and three scatterer-to-absorber distances d (10 cm, 15 cm, 20 cm). The reconstructed central slices were obtained at the 20$^{th}$ iteration.

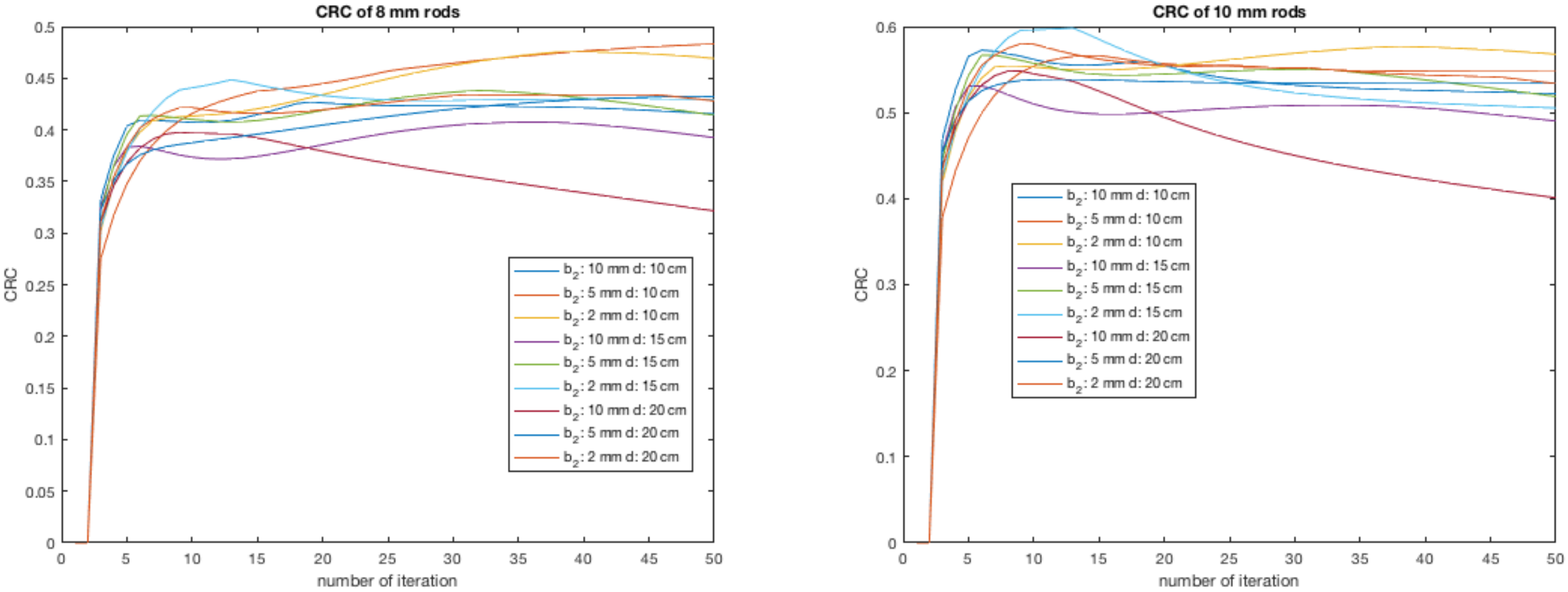

**Fig. 7.** Contrast recovery ratio (CRC) of the reconstructed 10 mm rods and 8 mm rods in figure 6.

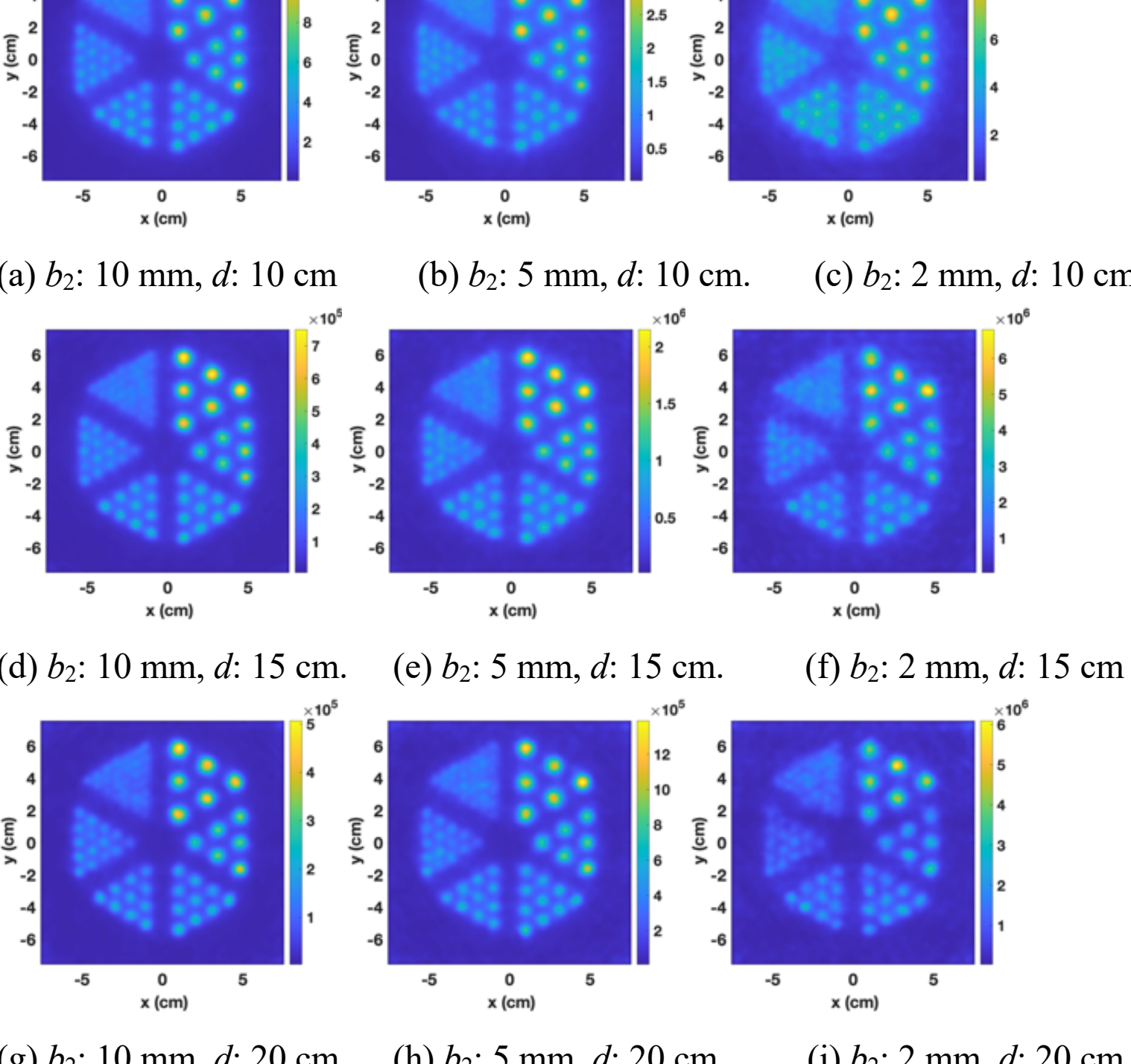

(a) $b_2$: 10 mm, *d*: 10 cm (b) $b_2$: 5 mm, *d*: 10 cm. (c) $b_2$: 2 mm, *d*: 10 cm

(d) $b_2$: 10 mm, *d*: 15 cm. (e) $b_2$: 5 mm, *d*: 15 cm. (f) $b_2$: 2 mm, *d*: 15 cm

(g) $b_2$: 10 mm, *d*: 20 cm. (h) $b_2$: 5 mm, *d*: 20 cm. (i) $b_2$: 2 mm, *d*: 20 cm

**Fig. 8.** Reconstruction results at 440 keV with source placed at 10 cm parallel to the detector plane. Simulations were conducted using three scatterer thickness $b_2$ 2 mm, 5 mm, 10 mm, and three scatterer-to-absorber distance $d$ (10 cm, 15 cm, 20 cm). The reconstructed central slices were obtained at the 20$^{th}$ iteration.

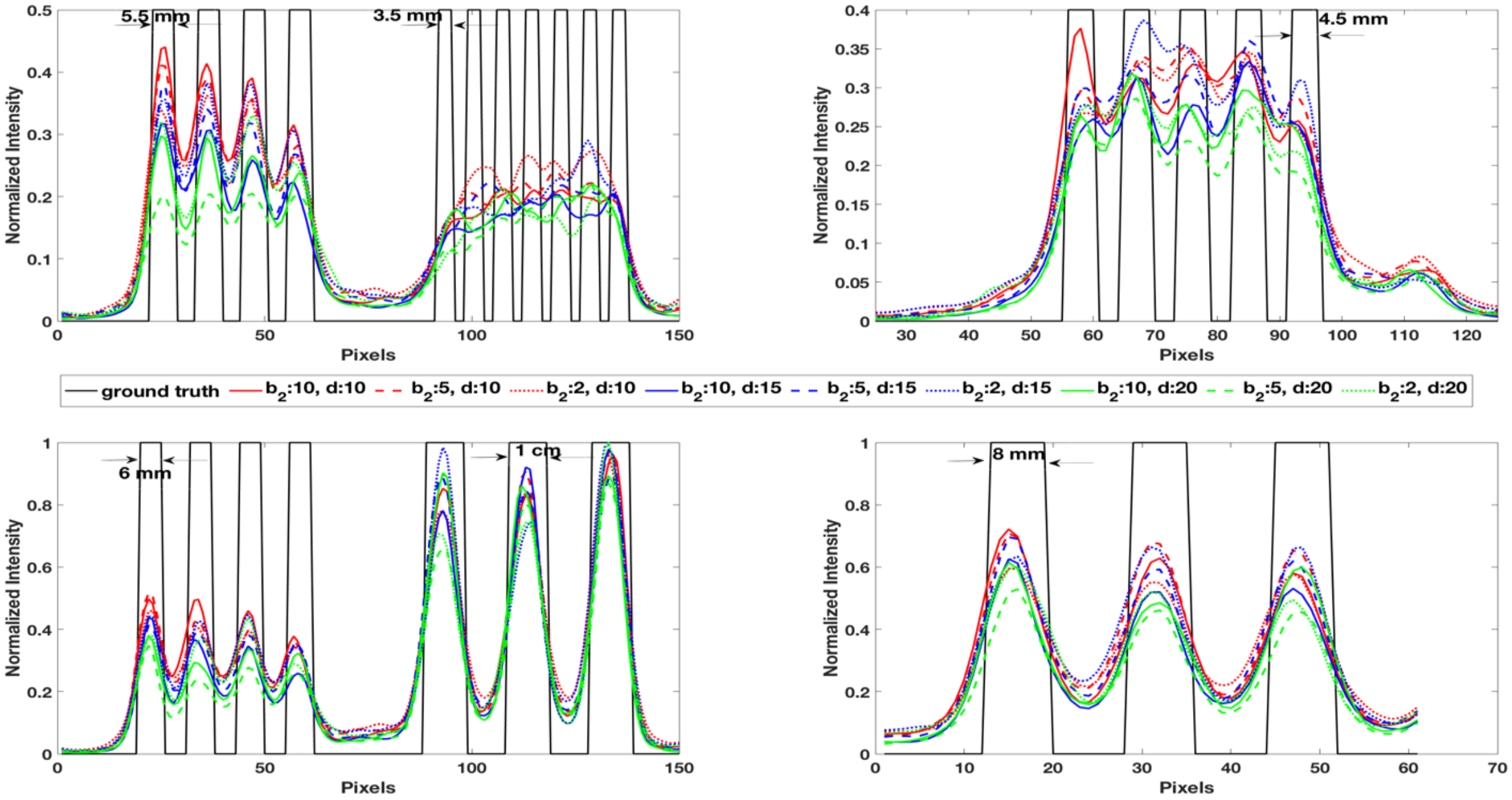

**Fig. 9.** Line profiles of the rods presented in figure 6, simulated with different scatterer thickness $b_2$ and different distances between absorber and scatterer d, the source is simulated at 218 keV.

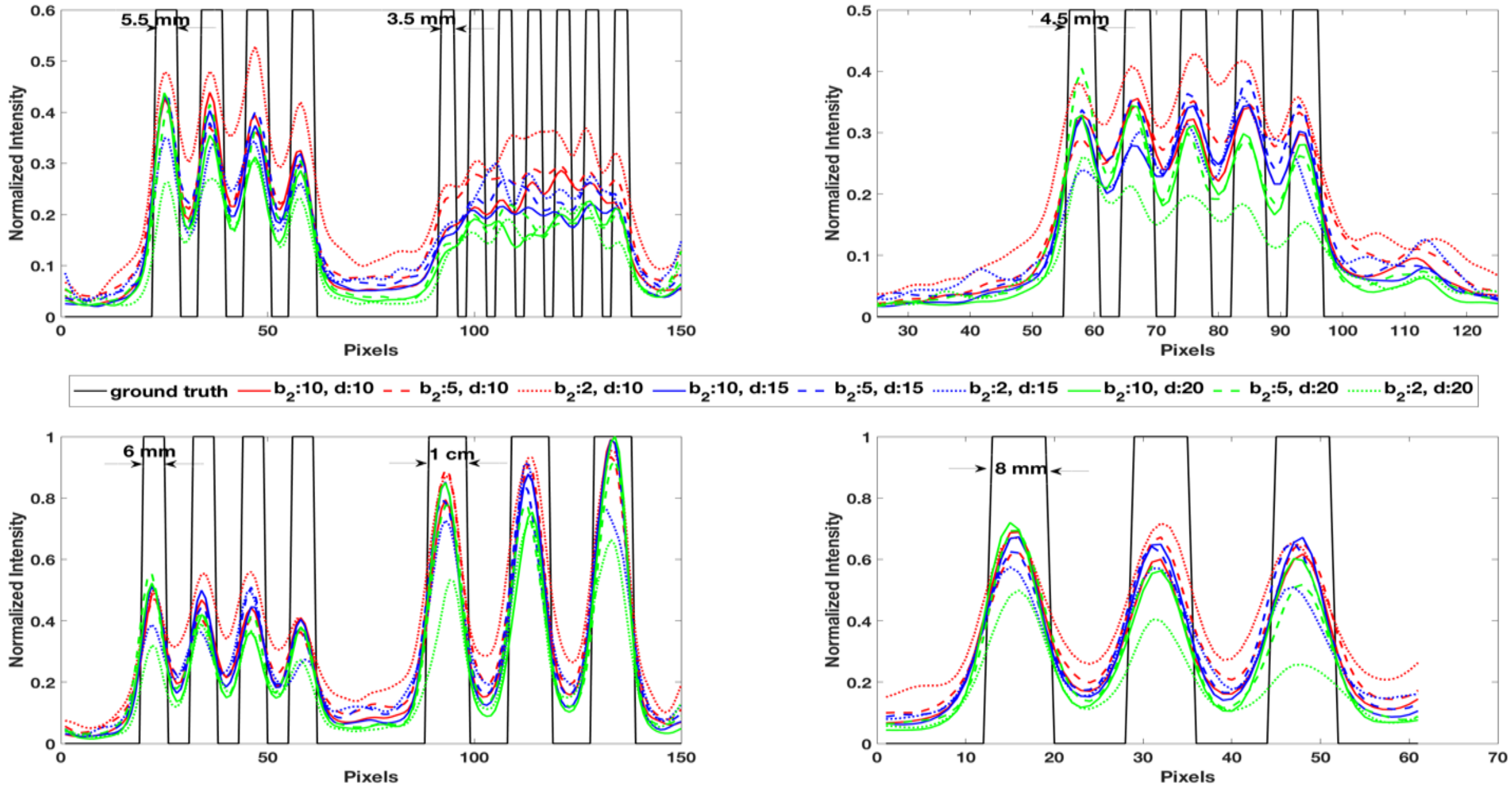

**Fig. 10.** Line profiles of the rods presented in figure 8, simulated with different scatterer thickness $b_2$ and different distances between absorber and scatterer d, the source is simulated at 440 keV.

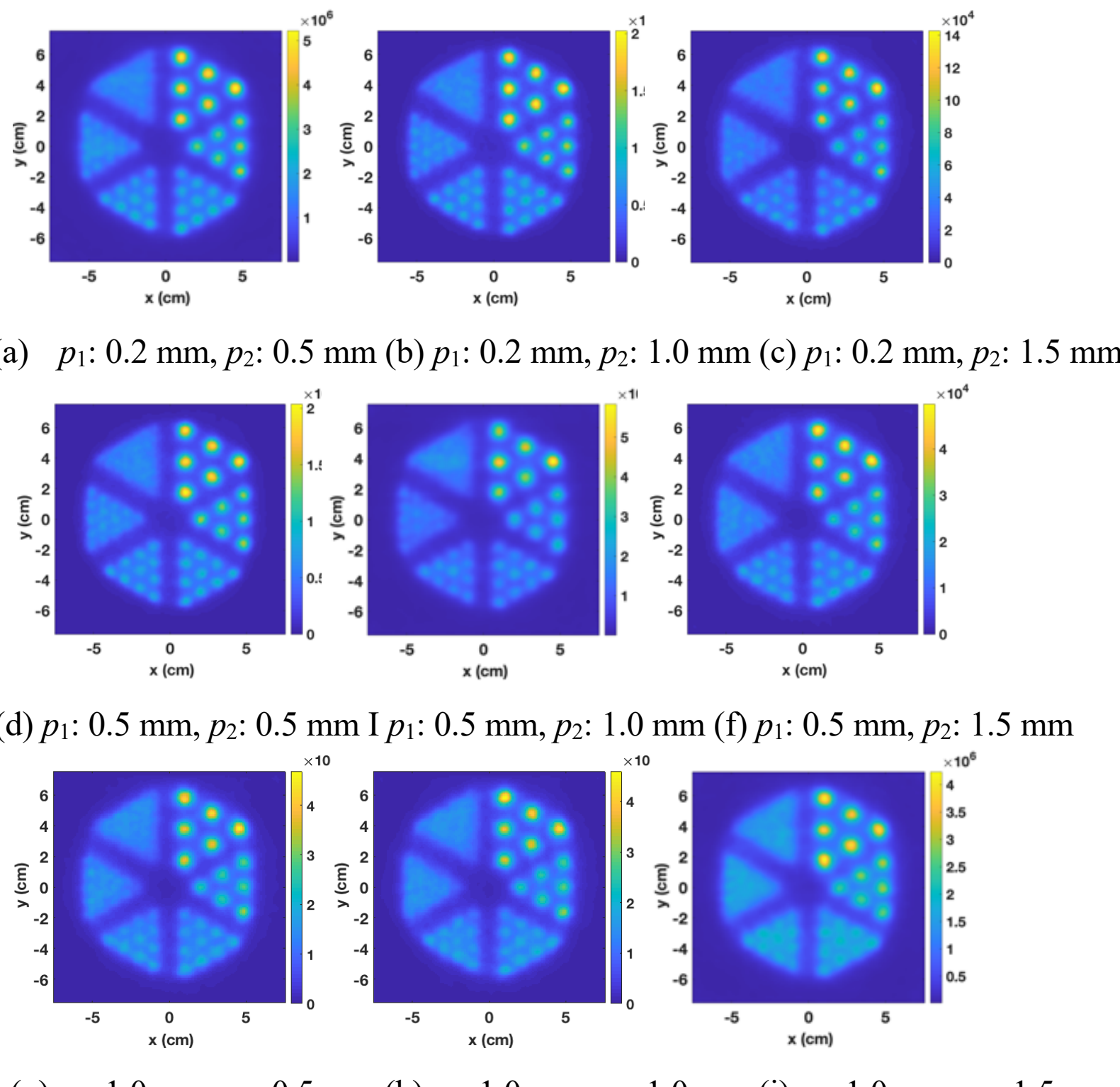

(a) $p_1$: 0.2 mm, $p_2$: 0.5 mm (b) $p_1$: 0.2 mm, $p_2$: 1.0 mm (c) $p_1$: 0.2 mm, $p_2$: 1.5 mm

(d) $p_1$: 0.5 mm, $p_2$: 0.5 mm I $p_1$: 0.5 mm, $p_2$: 1.0 mm (f) $p_1$: 0.5 mm, $p_2$: 1.5 mm

(g) $p_1$: 1.0 mm, $p_2$: 0.5 mm (h) $p_1$: 1.0 mm, $p_2$: 1.0 mm (i) $p_1$: 1.0 mm, $p_2$: 1.5 mm

**Fig. 11.** Reconstruction results at 218 keV with source placed at 10 cm parallel to the detector plane. Three scatterer pixel resolution $p_1$ 0.2 mm, 0.5 mm, 1 mm were simulated. Three absorber pixel resolution $p_2$ 0.5 mm, 1 mm, 1.5 mm were simulated. The reconstructed central slices were obtained at the 20$^{th}$ iteration. The scatterer thickness was 5 mm, the scatterer-to-absorber distance was 10 cm.

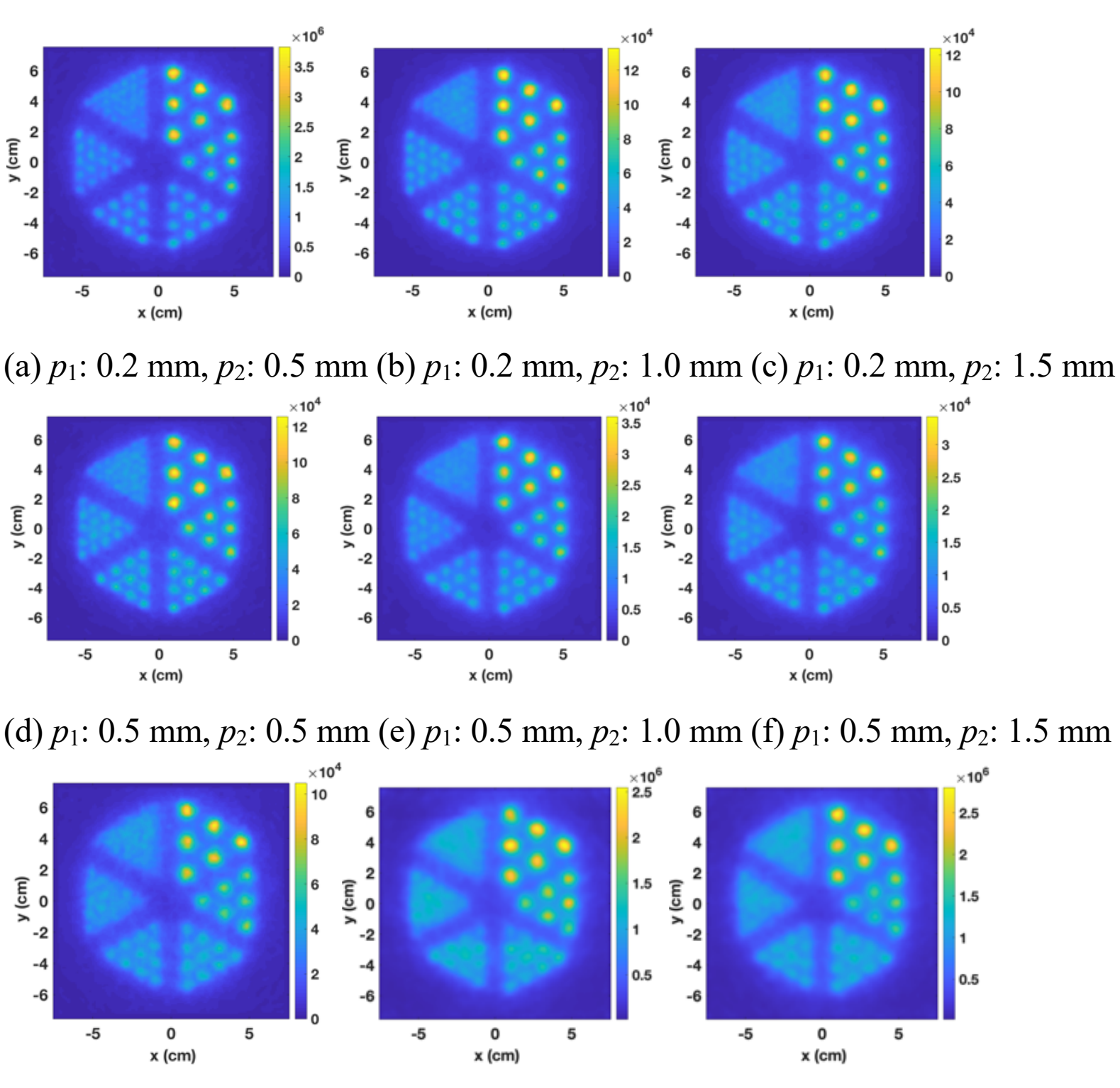

(a) $p_1$: 0.2 mm, $p_2$: 0.5 mm (b) $p_1$: 0.2 mm, $p_2$: 1.0 mm (c) $p_1$: 0.2 mm, $p_2$: 1.5 mm

(d) $p_1$: 0.5 mm, $p_2$: 0.5 mm (e) $p_1$: 0.5 mm, $p_2$: 1.0 mm (f) $p_1$: 0.5 mm, $p_2$: 1.5 mm

(g) $p_1$: 1.0 mm, $p_2$: 0.5 mm (h) $p_1$: 1.0 mm, $p_2$: 1.0 mm (i) $p_1$: 1.0 mm, $p_2$: 1.5 mm

**Fig. 12.** Reconstruction results at 440 keV with source placed at 10 cm parallel to the detector plane. Three scatterer pixel resolution $p_1$ 0.2 mm, 0.5 mm, 1 mm were simulated. Three absorber pixel resolution $p_2$ 0.5 mm, 1 mm, 1.5 mm were simulated. The reconstructed central slices were obtained at 20th iteration. The scatterer thickness was 5 mm, the scatterer-to-absorber distance was 10 cm.

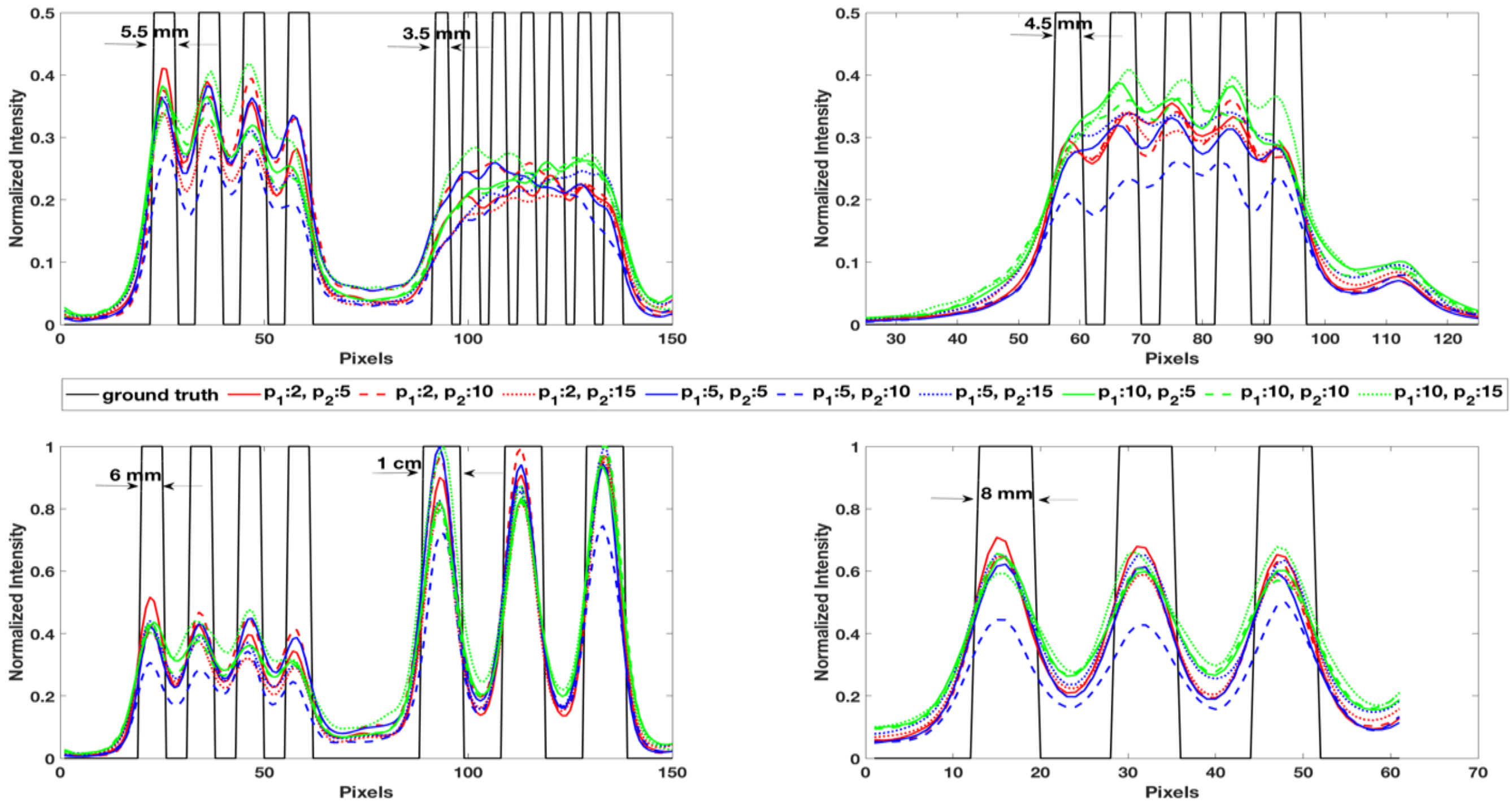

**Fig. 13.** Line profiles of the rods presented in figure 11, simulated with different pixels sizes of scatterer and absorber, the source is simulated at 218 keV.

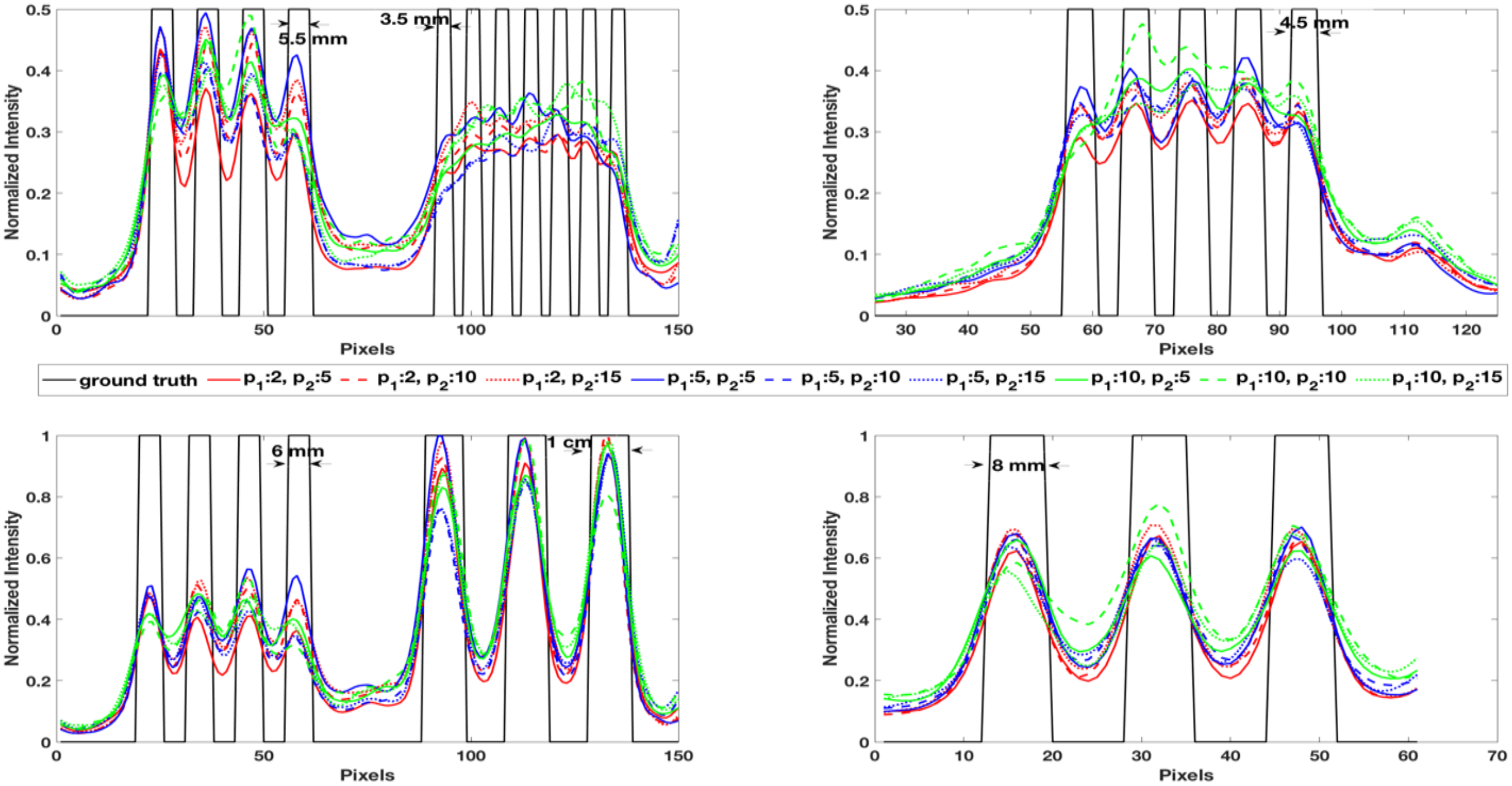

**Fig. 14.** Line profiles of the rods presented in figure 12, simulated with different pixels sizes of scatterer and absorber, the source is simulated at 440 keV.

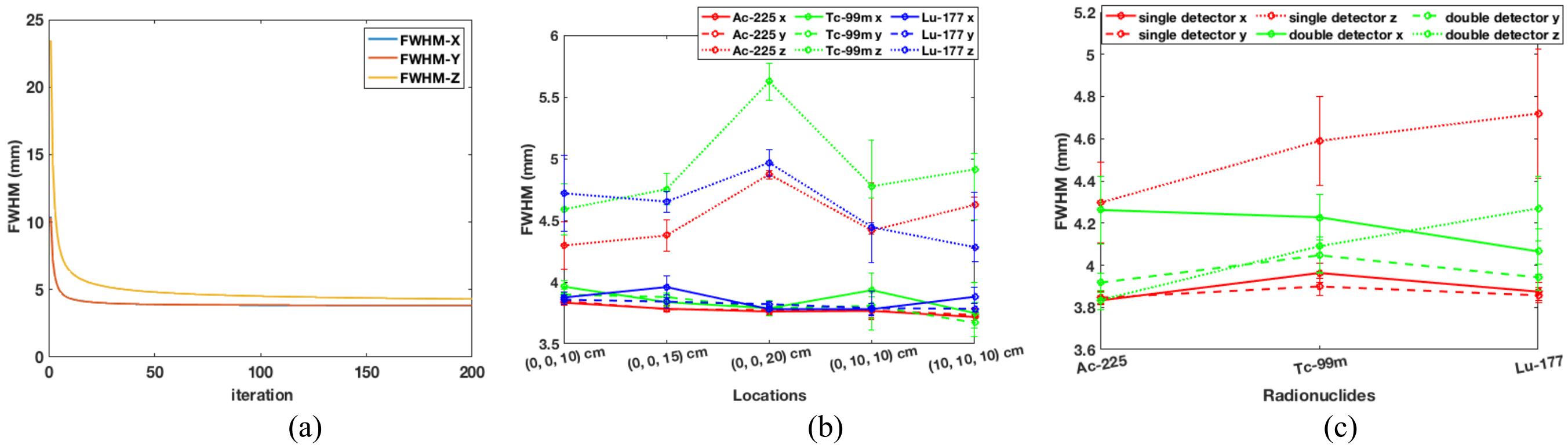

**Fig. 15.** (a): FWHM of the reconstructed point source along 3 dimensions with varying numbers of iterations, the point source filled with $^{225}Ac$ ($10^6$ primary particles) is simulated to be located at 10 cm distance from the single detector, **(**b): FWHMs of the reconstructed point sources placed at different locations and filled with $^{99m}Tc$, $^{117}Lu$, $^{225}Ac$ along 3 dimensions. The locations (x, y, z) cm are given with the center of the scattering layer of single detector set as (0, 0, 0), (c): comparison of the FWHMs obtained at 200$^{th}$ iteration of a reconstructed point, between single and double detectors.

To measure the FWHM of the reconstructed point-like source, we used the results at 200$^{th}$ iteration and applied a Gaussian fit along 3 dimensions. The 200$^{th}$ iteration was chosen because the FWHM of the reconstructed point tends to stabilize, as evidenced in figure 15 (a), which presents the FWHM of the fitted Gaussian as a function of the number of iterations for a point-like source containing $^{225}Ac$, located at 10 cm from the center of the scattering layer without any shift parallel to the detector plane. Figure 15 (b) demonstrates the FWHMs of resolution at 200$^{th}$ iteration at different locations with different radionuclides simulated. The error bars are added based on the uncertainties of the fitted Gaussians in MATLAB.

The FWHM in the plane parallel to the detector is less than 4.0 mm for all the tested sources, and for $^{225}Ac$, the FWHM is as low as 3.7 mm. For the source filled with $^{99m}Tc$, which is expected to be less advantageous in Compton imaging mode compared to conventional SPECT due to the lower gamma energy, the FWHM in the vertical direction is 5.6 mm, and 3.9 mm in the plane parallel to the detector.

TABLE VII

PROPORTION OF MULTI-SCATTERING VS. FULL ABSORPTION IN 2 INTERACTIONS-EVENTS WITH INITIAL ENERGY OF 440 KEV

| | |
|---|---|
| $E_0 \in$[430, 450] keV | 9% |
| Multi-scattering | 91% |

TABLE VIII

PROPORTION OF MULTI-SCATTERING VS. FULL ABSORPTION IN 2 INTERACTIONS-EVENTS WITH INITIAL ENERGY OF 218 KEV

| | |
|---|---|
| $E_0 \in$[211, 225] keV | 28% |
| Multi-scattering | 71% |

TABLE IX

COUNTS USED IN THE RECONSTRUCTION FOR DIFFERENT GEOMETRY PARAMETERS: DETECTOR DISTANCE D AND SCATTERER THICKNESS B2, 218 KEV

| $b_2$ \ d | 10 mm | 5 mm | 2 mm |
|---|---|---|---|
| 10 cm | $1.26\times10^6$ | $2.01\times10^6$ | $1.76\times10^6$ |
| 15 cm | $8.51\times10^5$ | $1.30\times10^6$ | $1.08\times10^6$ |
| 20 cm | $5.96\times10^5$ | $8.87\times10^5$ | $7.08\times10^5$ |

TABLE X

COUNTS USED IN THE RECONSTRUCTION FOR DIFFERENT GEOMETRY PARAMETERS: DETECTOR DISTANCE D AND SCATTERER THICKNESS B2, 440 KEV

| $b_2$ \ d | 10 mm | 5 mm | 2 mm |
|---|---|---|---|
| 10 cm | $1.84\times10^6$ | $1.70\times10^6$ | $1.06\times10^6$ |
| 15 cm | $1.04\times10^6$ | $9.15\times10^5$ | $5.47\times10^5$ |
| 20 cm | $6.49\times10^5$ | $5.48\times10^5$ | $3.13\times10^5$ |

TABLE XI

RESOLVED DIAMETER WITH DIFFERENT SCATTERER PIXEL SIZE P1, ABSORBER PIXEL SIZE P2 SIMULATED, SCATTERER THICKNESS IS 5 MM, 218 KEV.

| $P_2$ \ $P_1$ | 0.5 mm | 1.0 mm | 1.5 mm |
|---|---|---|---|
| 0.2 mm | 4.5 mm | 4.5 mm | 4.5 mm |
| 0.5 mm | 4.5 mm | 4.5 mm | 4.5 mm |
| 1.0 mm | 5.5 mm | 5.5 mm | 5.5 mm |

TABLE XII

RESOLVED DIAMETER WITH DIFFERENT SCATTERER PIXEL SIZE $P_1$, ABSORBER PIXEL SIZE $P_2$ SIMULATED, SCATTERER THICKNESS IS 5 MM, 440 KEV.

| $P_2$ / $P_1$ | 0.5 mm | 1.0 mm | 1.5 mm |
| --- | --- | --- | --- |
| 0.2 mm | 3.5 mm | 3.5 mm | 3.5 mm |
| 0.5 mm | 3.5 mm | 4.5 mm | 5.5 mm |
| 1.0 mm | 5.5 mm | 5.5 mm | 5.5 mm |

*B. Reconstruction of two detectors, fixed mode*

The FWHM of the reconstructed point-like source with two detectors simulated is shown in figure 15 (c), measured at 200th iteration. The error bars are added using the uncertainties of fitted Gaussians. The results are compared to single detector. When the second detector is added, positioned at the edge of the first detector, the FWHM in the z direction is improved, from 4.3 mm to 3.8 mm for $^{225}Ac$, from 4.6 mm to 4.2 mm for $^{99m}Tc$, and from 4.7 mm to 4.2 mm for $^{177}Lu$.

Reconstructed results of the Derenzo phantom with 8 cm length at 218 keV and 440 keV are shown in figure 16 and figure 17. The reconstructed results with a mixture of 218 keV and 440 keV are shown in figure 18. In the reconstruction, only the photons undergoing interactions at different layers and having total energy deposited within windows (211-225 keV, 430-450 keV) are used.

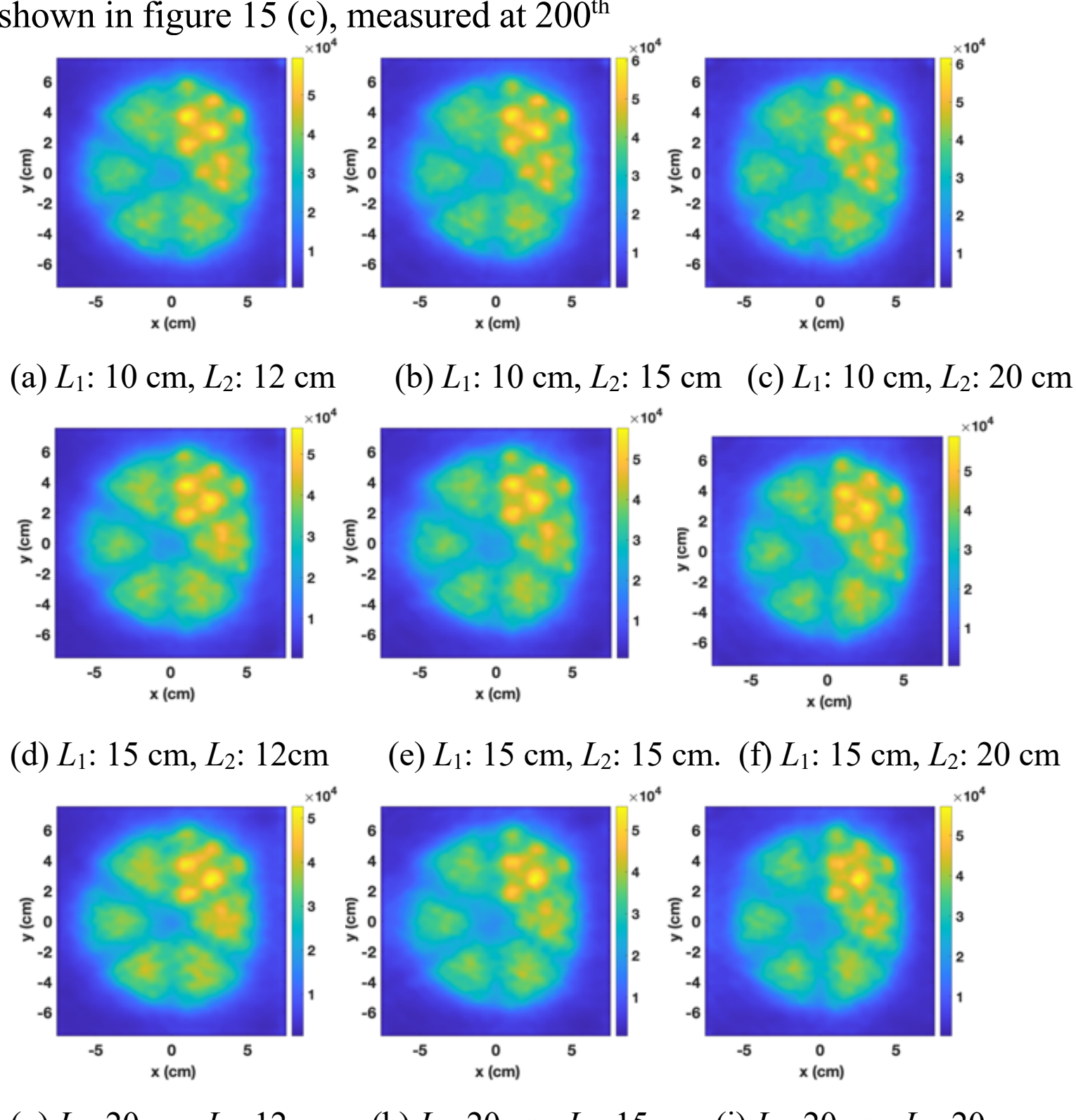

(a) $L_1$: 10 cm, $L_2$: 12 cm (b) $L_1$: 10 cm, $L_2$: 15 cm (c) $L_1$: 10 cm, $L_2$: 20 cm

(d) $L_1$: 15 cm, $L_2$: 12cm (e) $L_1$: 15 cm, $L_2$: 15 cm. (f) $L_1$: 15 cm, $L_2$: 20 cm

(g) $L_1$: 20 cm, $L_2$: 12 cm (h) $L_1$: 20 cm, $L_2$: 15 cm. (i) $L_1$: 20 cm, $L_2$: 20 cm

**Fig. 16.** Reconstruction results at 218 keV.

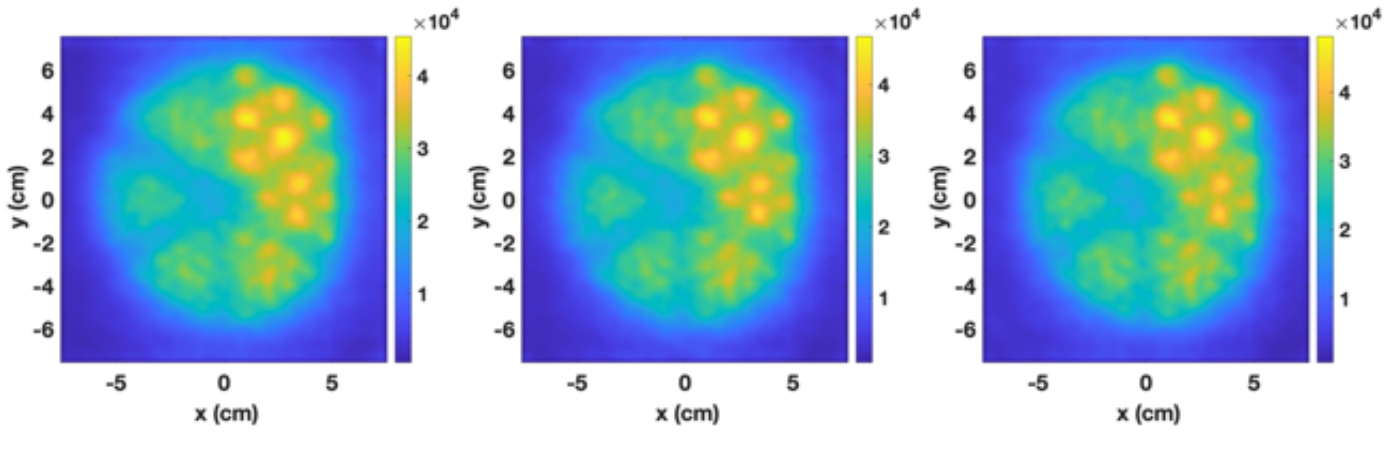

(a) $L_1$: 10 cm, $L_2$: 12 cm (b) $L_1$: 10 cm, $L_2$: 15 cm (c) $L_1$: 10 cm, $L_2$: 20 cm

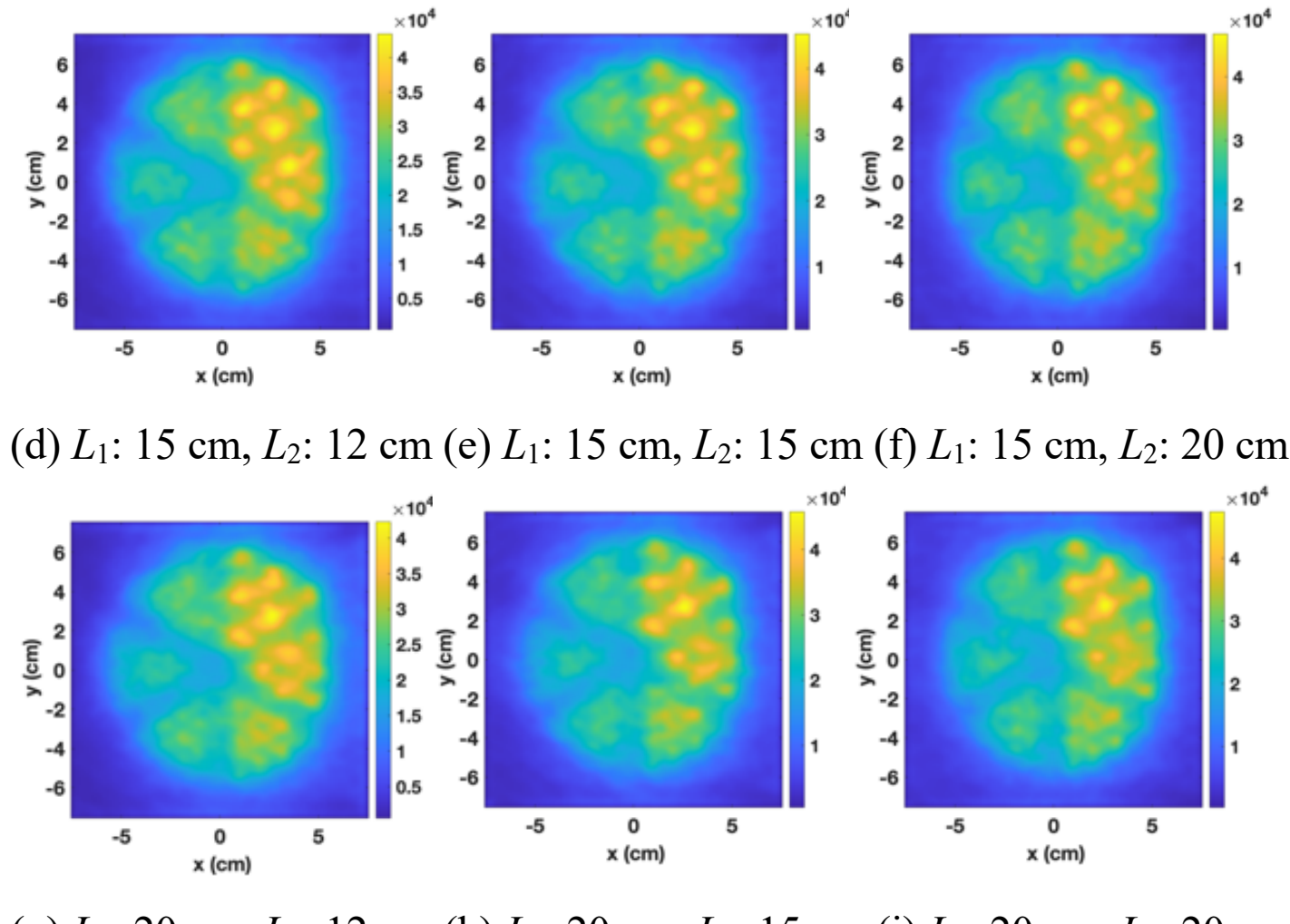

(d) $L_1$: 15 cm, $L_2$: 12 cm (e) $L_1$: 15 cm, $L_2$: 15 cm (f) $L_1$: 15 cm, $L_2$: 20 cm

(g) $L_1$: 20 cm, $L_2$: 12 cm (h) $L_1$: 20 cm, $L_2$: 15 cm (i) $L_1$: 20 cm, $L_2$: 20 cm

**Fig. 17.** Reconstruction results at 440 keV.

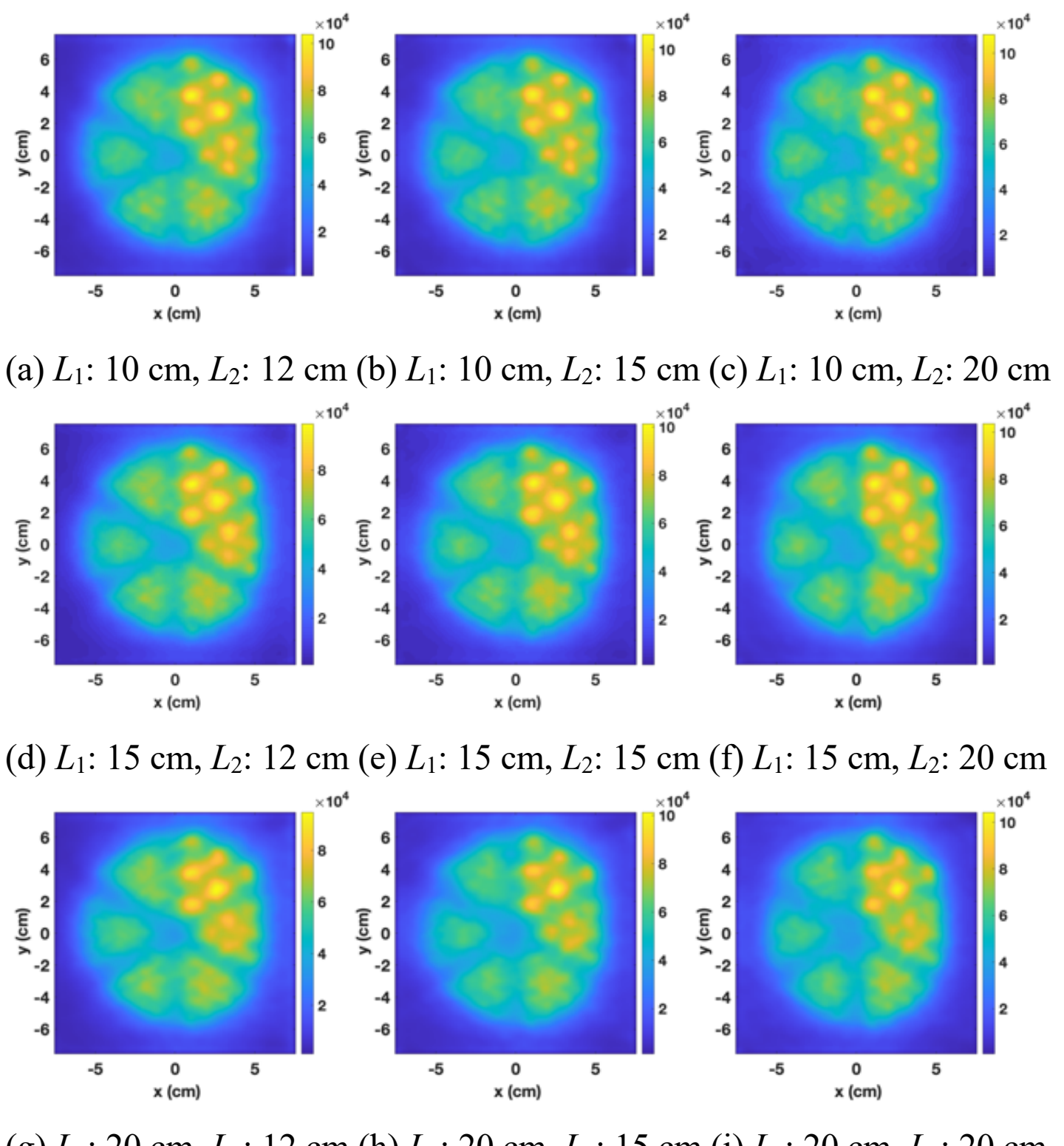

(a) $L_1$: 10 cm, $L_2$: 12 cm (b) $L_1$: 10 cm, $L_2$: 15 cm (c) $L_1$: 10 cm, $L_2$: 20 cm

(d) $L_1$: 15 cm, $L_2$: 12 cm (e) $L_1$: 15 cm, $L_2$: 15 cm (f) $L_1$: 15 cm, $L_2$: 20 cm

(g) $L_1$: 20 cm, $L_2$: 12 cm (h) $L_1$: 20 cm, $L_2$: 15 cm (i) $L_1$: 20 cm, $L_2$: 20 cm

**Fig. 18.** Reconstruction results with 218 keV and 440 keV reconstructed simultaneously.

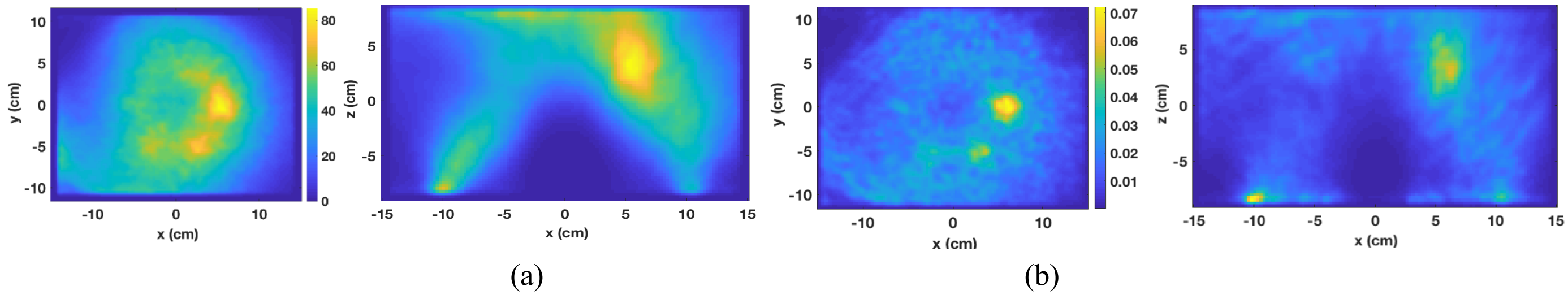

(a) (b)

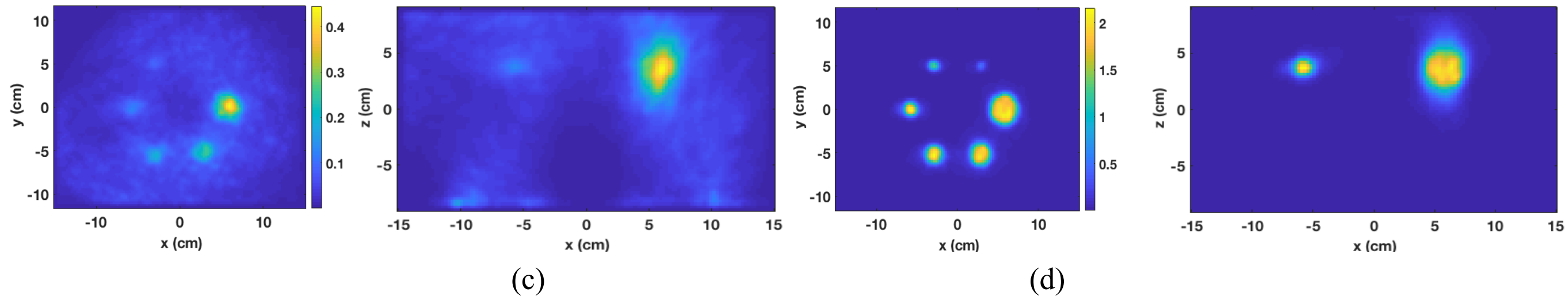

**Fig. 19.** Reconstruction results of the simulated NEMA phantom: (a) filled with 350 MBq $^{99m}$Tc, (b) filled with 5.7 MBq $^{225}$Ac, target to background ratio of 12:1, (c) 5.7 MBq $^{225}$Ac, target to background ratio of 30:1, (d) 5.7 MBq $^{225}$Ac in cold background. Images are shown at the 50$^{th}$ iteration of list mode MLEM. Slices are displayed in the center of the transverse and coronal planes.

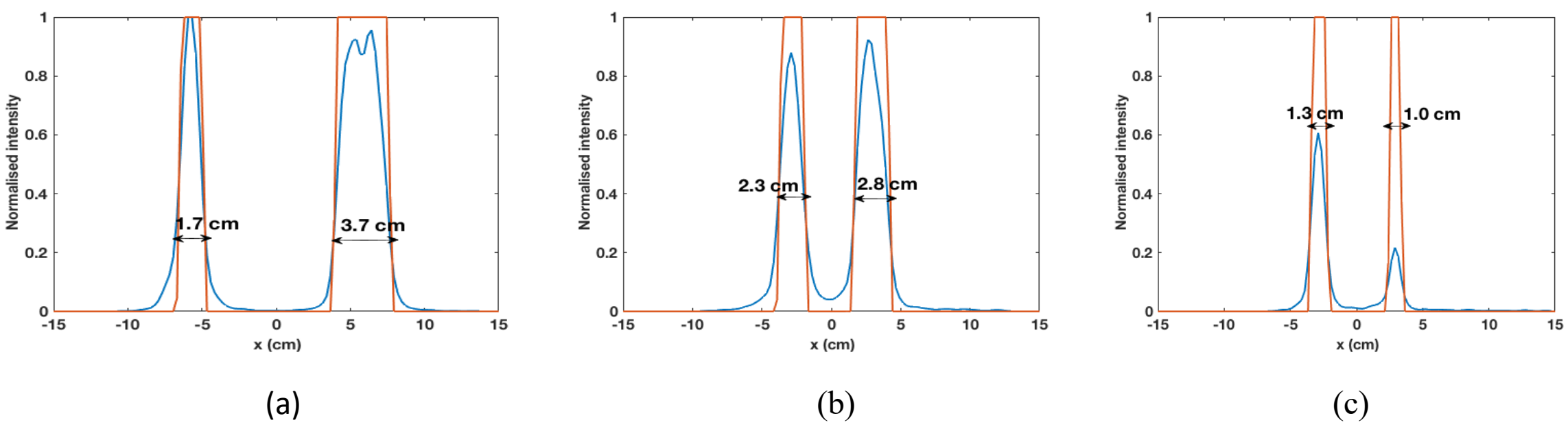

**Fig. 20.** Central line profiles of the reconstructed spheres in NEMA IQ phantom in the transverse plane, the source filled with $^{225}$Ac in water was simulated with cold background.

The central slices of the transverse plane and coronal plane of the reconstructed NEMA IQ phantom are shown in figure 19. We draw line profiles of each sphere simulated with a cold background in figure 20. The calculated RC and CRC are summarized in table 13-15. The accuracies are calculated using the standard deviation of the hot rods and shown in tables. For the reconstruction of the NEMA IQ phantom, the same energy windows used for the point-like source are applied for event selection, 211-225 keV and 430-450 keV for $^{225}$Ac, and 135.5-147.5 keV for $^{99m}$Tc.

TABLE XIII. CONTRAST RECOVERY OF $^{99M}$TC, HOT RATIO 8:1

| diameter (cm) metric(%) | 1.0 | 1.3 | 1.7 | 2.2 | 2.8 | 3.7 |
|---|---|---|---|---|---|---|
| RC | 6.000 ± 0.027 | 6.000 ± 0.018 | 5.000 ± 0.015 | 7.000 ± 0.004 | 9.000 ± 0.005 | 10.000 ± 0.004 |
| CRC | 11.000 ± 0.095 | 7.000 ± 0.066 | 4.000 ± 0.054 | 12.000 ± 0.015 | 21.000 ± 0.017 | 25.000 ± 0.013 |

TABLE XIV. CONTRAST RECOVERY COEFFICIENT (%) OF $^{225}$AC

| diameter (cm) hot ratio | 1.0 | 1.3 | 1.7 | 2.2 | 2.8 | 3.7 |
|---|---|---|---|---|---|---|
| 12:1 | 4.000 ± 0.271 | 4.000 ± 0.094 | 6.000 ± 0.089 | 10.000 ± 0.048 | 15.000 ± 0.040 | 21.000 ± 0.037 |
| 30:1 | 4.000 ± 0.256 | 11.000 ± 0.217 | 19.000 ± 0.150 | 28.000 ± 0.201 | 39.000 ± 0.148 | 56.000 ± 0.121 |

TABLE XV. RECOVERY COEFFICIENT (%) OF $^{225}$AC

| diameter (cm) hot ratio | 1.0 | 1.3 | 1.7 | 2.2 | 2.8 | 3.7 |
|---|---|---|---|---|---|---|
| 12:1 | 2.000 ± 0.050 | 2.000 ± 0.017 | 2.000 ± 0.016 | 3.000 ± 0.009 | 4.000 ± 0.007 | 6.000 ± 0.006 |
| 30:1 | 5.000 ± 0.002 | 8.000 ± 0.002 | 10.000 ± 0.001 | 13.000 ± 0.002 | 17.000 ± 0.001 | 23.000 ± 0.001 |
| Cold background | 7.000 ± 0.005 | 19.000 ± 0.009 | 31.000 ± 0.008 | 32.000 ± 0.005 | 32.000 ± 0.002 | 34.000 ± 0.001 |

## IV. DISCUSSION

The impact of the detector parameters on the reconstructed image resolution was investigated in this study. For the single detector, several parameters were simulated, including scatterer-absorber distance, scatterer thickness, and scatterer and absorber pixel sizes. The optimized parameters were selected based on achieving the best possible imaging resolution with the current system matrix modeling used in the reconstruction. It was observed that a larger detector pixel size results in reduced imaging resolution as shown in figure 11 and 12. Selecting the scatterer thickness requires a delicate balance. A smaller thickness results in fewer detected photons, whereas a larger thickness can stop the photons, leading to multiple interactions on the same scatterer that generate false angles in the reconstruction. A scatterer thickness of 5 mm was chosen based on Monte Carlo simulation. For the two-detector system designed for human whole-body imaging, based on the results obtained from a single detector, the following parameters were chosen: a scatterer-to-absorber distance of 10 cm, a scatterer thickness of 5 mm, a scatter pixel size of 0.2 mm, and an absorber pixel size of 0.5 mm. All image reconstructions were performed using the system matrix from [9] and the sensitivity matrix from [37]. Although previous studies have demonstrated agreement between the calculated sensitivity and simulated sensitivity, a comparison of the system matrix with the simulation of alpha particles is still required. We believe that image resolution could be further improved by employing a more accurate system matrix. The current model considers the solid angle, conical uncertainties, and the Klein-Nishina cross section. We will incorporate the possibility of absorption within scatterer, along with a more precise Doppler broadening model. We will implement a Gaussian function with a standard deviation that varies with the detected energies for the Doppler broadening effects in the absorber. We will also explore other functions for modeling Doppler broadening, such as the Cauchy distribution, and compare them with Monte Carlo simulated data in future study.

Despite the image reconstruction methods, several other factors affect the quality of reconstructed images when using a Compton camera for alpha imaging. These factors include system geometry design, readout electronics, Doppler broadening, and the detected counts due to low injected dose. Doppler broadening was modeled in the simulation, and the results at 218 keV, shown in figure 6 (a) and figure 11 (a), exhibit lower image quality compared to the 440 keV shown in figure 8 (a) and figure 12 (a). This is due to the Doppler effects and larger angular uncertainties caused by greater energies uncertainties at lower energy. The difference in reconstructed images between the two energy peaks is more evident in the 2D Derenzo phantom, where the dose concentration is as high as 0.84 MBq/cc. This high dose concentration minimizes the influence of limited counts compared to a concentration of 0.1 MBq/cc. For imaging the 0.1 MBq/cc Derenzo phantom with an 8 cm length, there are no noticeable differences between the results of two energy peaks shown in figure 16 and figure 17, as the impacts of the detector's intrinsic resolution and low detected counts are more significant than the Doppler broadening effects.

We selected the optimized geometry parameters of a single detector, and we separately simulated point-like sources filled with three radionuclides ($^{99m}$Tc, $^{177}$Lu, $^{225}$Ac) at different locations. The entire decay chain of these three radionuclides is included in the simulation. The sources were placed in a water phantom with a volume as small as the point-like source. The goal of simulating the point-like source was to evaluate the FWHM of the resolution in a noise-free background of the FOV. Figure 15 (b) shows the FWHMs of the reconstructed point-like sources located at different positions, using the center of the first scattering layer as the reference point. For a single detector, the FWHM in the plane parallel to the detector surface is better than in the direction perpendicular to the detector surface. This is due to the larger uncertainties in determining the conical surface intersection in the vertical direction. Among the three tested radionuclides, results of $^{225}$Ac outperform the others, with an FWHM smaller than 5.0 mm for a source placed at a distance of 20 cm from the detector.

In figure 15 (c), the double detector mode is compared to the single detector by measuring the FWHM of the reconstructed point source located 10 cm from each detector. This simulation demonstrates the improvements the second detector can provide for image resolution in the vertical direction of the patient bed. However, the dimension parallel to the detector surface is not improved and is slightly disturbed by the second detector, as the vertical direction of the second detector with larger uncertainties is added to this plane.

In our study, for evaluating the performance of our whole-body system, a Derenzo phantom with pure gamma emission in air and a NEMA IQ phantom filled with $^{225}$Ac in water with cold and warm background are simulated. We simulated a Derenzo phantom with activity concentration of 0.1 MBq/cc. With the simplified simulation setting, where only the two gamma photons with energies of 218 keV and 440 keV are simulated, we demonstrate that an achievable resolution of 1 cm is possible when the source is placed close to the detector (figure 18 (a)). However, when the source is placed 20 cm away, the resolution degrades, and the 1 cm rods become difficult to distinguish. For the NEMA IQ phantom, the full decay chain of $^{225}$Ac is simulated, and energy windows (211-225 keV, 430-450 keV) are applied for event selection in the reconstruction. The results with a cold background suggest a resolution of 1 cm with 5.7 MBq of $^{225}$Ac simulated with an acquisition time of 15 minutes. The contrast recovery decreases with a warm background. For imaging the NEMA phantom filled with $^{225}$Ac with a hot to background activity ratio 12:1, only the 3.7 cm sphere can be resolved, and with a hot to background ratio 30:1, the 1.3 cm sphere is resolved.

We have simulated spheres of the same diameters (3.7 cm, 2.8 cm, 2.2 cm, 1.7 cm, 1.3 cm and 1 cm) as in [22], and with an activity concentration 30 kBq/ml and a 15 minutes acquisition in the water background, the 1.0 cm sphere is resolved, as shown in figure 19 (d). This indicates that with a scanning time comparable to conventional SPECT, and using the same dose level of $^{225}$Ac simulated in a cold water background, the simulated Compton camera system achieves a resolution of 1.0 cm for human whole-body imaging, as measured with a NEMA IQ phantom using the current reconstruction method. This resolution surpasses that of current commercial SPECT scanners typically used for whole-body applications for imaging $^{225}$Ac ([22]), where the smallest resolvable sphere has a diameter of 1.3 cm with an injected concentration of 35 kBq/ml of $^{225}$Ac in the NEMA phantom. The evaluated systems in [22] include the Optima 640 (GE), the Discovery 670 (GE) with a HEGP collimator, and the Symbia T6 (Siemens). Among them, only the Discovery 670 (GE) could resolve a 1.3 cm sphere, while the Optima 640 (GE) with a MEGP collimator could only resolve a 2.2 cm sphere. The improved 1.0 cm resolution achieved by the Compton camera does not represent a dramatic breakthrough in imaging resolution but does offer an improvement over the commercial SPECT systems. This result demonstrates the advantages that a Compton camera can bring for imaging higher energetic gamma photons, especially in targeted alpha therapy using $^{225}$Ac, which has limited gamma emission.

However, for imaging the lower energy sources such as $^{99m}$Tc, conventional SPECT systems generally achieve a resolution of 12-16 mm [14]. In our tests, the Compton camera system demonstrated limited advantages in this context. Compared to novel CZT-based collimated SPECT systems ([16, 17] [15]) and conventional anger camera, Compton camera imaging provides lower image quality when photon energies are in the lower energy range, due to the increasing Doppler broadening effects and reduced possibility of Compton scattering at lower energies. We simulated a NEMA IQ phantom filled with 320 MBq of $^{99m}$Tc. With a 10-minute acquisition, only the 3.7 cm and 2.8 cm spheres could be resolved, and the contrast recovery ratio was approximately 10%. In comparison, a study on the CZT-based SPECT demonstrated a resolution better than 9.5 mm for cold rod recovery in a Derenzo phantom filled with $^{99m}$Tc and background ([15]), with the same acquisition time 10 minutes, this indicates that the performance of these systems surpasses that of the presented Compton camera system. The FWHM obtained with a point-source filled with $^{99m}$Tc, simulated at a 10 cm distance from the detector is ~4 mm (figure 15 (c)), in contrast, for the CZT-based SPECT [15], the achievable FWHM with the same radionuclide at the center of the FOV is 3.54 mm, and for an anger camera it is even better (2.47 mm). The quantification accuracy of $^{177}$Lu, a beta emitter, is presented in [16, 17], showing that the image quality obtained with CZT-based SPECT outperforms conventional SPECT. It would be valuable to compare the Compton camera system with the CZT SPECT system for $^{177}$Lu imaging, as this is the most used beta emitters in TRT. Since the goal of this study is to explore Compton camera for $^{225}$Ac, we reserve the imaging of $^{177}$Lu for future work.

This work has several limitations. The first is that, for the radionuclide $^{225}$Ac, only two energy peaks are simulated and reconstructed for a Derenzo phantom. Other photon emitters such as $^{213}$Po and $^{217}$At are not simulated in the Derenzo phantom, although they would contribute to random detections. A more comprehensive simulation of gamma emitters, as well as random and scatter events, should be conducted to further evaluate the imaging performance of the proposed system. Additionally, the two energy peaks were simulated separately in Derenzo phantom. Down-scattering caused by 440 keV scattered photons were not simulated and will be investigated in future work.

Only $^{225}$Ac is used in this study, and the gamma energy generated in TAT can range from 60 keV to 2.6 MeV, depending on the alpha emitters. Higher energy photons will introduce energy misclassification during the detection of gamma rays. We only evaluated $^{225}$Ac because it is the most often investigated agent for TAT, but more investigation is needed to assess the capability of imaging multi-energetic gamma rays generated from other existing alpha emitters, such as $^{211}$At or $^{227}$Th. For the NEMA phantom, the background environment was simulated as water. In contrast, evaluating a whole-body system requires modeling bone, fat, and blood in the phantom. Using a 4D-XCAT phantom as introduced in [46] involves specific design considerations for the dose, such as lesion location, lesion size, organ consumption, and dose administered to the lesion. A 3D printed model of a human head based on the CT scan was created and used for evaluating the quantification of $^{225}$Ac imaging in glioblastomas in [47]. A realistic tumor phantom is implemented and tested for validating the acquisition and reconstruction protocol in [22]. Currently, studies related to body phantom simulations with $^{225}$Ac are very limited, therefore, we chose to use only a NEMA IQ phantom with a size equivalent to a whole-body FOV to evaluate the performance of the system. However, evaluating the whole-body imaging performance using a patient phantom should be considered. Patient cases reported using $^{225}$Ac for the treatment of neuroendocrine tumor (NET), such as in [26] and [25], might be useful for designing a patient body phantom in the future.

Another limitation is the event selection methods. We select only the coincidences that involve interactions on different layers of detector and where the sum of the deposited energy from the incident gamma photon falls within an energy window. This selection limits the counts used in the reconstruction, which in turn diminishes one of the advantages of the Compton camera: its higher sensitivity compared to collimated SPECT. The statistics presented in table 7 indicate that for 440 keV gamma photons, the percentage of multi-scattering events is higher than for 218 keV gamma. It is evident that the likelihood of multiple scattering increases as photon energies increase. To increase the usable counts in the reconstruction, new algorithms including an event selection model and a new system matrix model adapted to multi-interaction coincidences will be developed in future work. Multi-scattering layers should also be simulated and compared to the current single scatterer system to increase sensitivity. Incorporating multiple scattered events into image reconstruction is challenging. The probability of multi-scattering should be incorporated with multi-scattered events included, and the energy uncertainty model should account for multiple uncertainties during different scatterings. We will implement methods that have been validated in proton therapy [48, 49]. Spectrum deconvolution should be considered to reduce the noises of the input. Machine learning methods, such as neural networks, might help increase the signal to background ratio ([50, 51]). Moreover, for $^{225}$Ac, the two major gamma rays (218 keV, 440 keV) generated from $^{221}$Fr and $^{213}$Bi are easy to separate, the system's performance needs to be investigated, especially for gammas having close energy levels. In reality backscattering events will contribute to blurring in the reconstructed images. The addition of TOF information might enhance resolution by allowing the filtering of backscattering events. Furthermore, the current system has a planar shape, varying distances between the scatterer pixels and absorber pixels, or even a spherical-shaped detector, may improve the imaging resolution and sensitivity, which requires further geometric design optimization. In the simulations, no DOI information is evaluated, the measurement uncertainties of detected locations are simulated in three dimensions to create a realistic scenario. In the reconstruction, the detected position in depth with measurement uncertainties equal to the pixel size is used. The impacts of the DOI information should be further investigated, as it is important for improving the quality of reconstruction. We have simulated CZT for the detector material. The atomic number Z of CZT is 50, which is higher than that of germanium (Z=32) and silicon (Z=14), two materials that are often chosen for scatterer due to their lower atomic numbers, reduced Doppler broadening effects and good energy resolution. However, Ge requires temperature cooling, Si has lower stopping power for gamma rays and lower detection efficiency, and their energy resolution and count rate are inferior to those of CZT. In the current state of art, the best achievable pixel size in a CZT detector is submillimeter [52]. But the current practical dimensions are ~1.5 mm, as seen in detector module M1085 from Redlen Technologies. The chosen pixel size in this work is feasible, but the cost of detector fabrication and readout electronics may be high.

While Compton cameras excel in imaging high-energy gamma rays, their primary application in human imaging seems to be in proton therapy monitoring. Nevertheless, our study indicates that for imaging of $^{225}$Ac, Compton cameras offer advantages in resolution recovery and greater sensitivity compared to collimated SPECT. We believe that the achievable imaging resolution for lower doses can be improved through refined event selection and modeling in the reconstruction process, by using the multi-scattering events.

## V. Conclusion

We demonstrated the feasibility of Compton camera imaging for whole-body applications for TAT treatment using Monte Carlo simulations. With a cold background and 5.7 MBq of $^{225}$Ac simulated, the 1 cm sphere in NEMA IQ phantom can be resolved with 15 minutes acquisition. In the presence of a warm background, with a 12:1 hot-to-warm ratio and hot concentration 1.6 kBq/ml, the 3.7 cm sphere can be resolved. Although the imaging resolution of the whole-body system does not surpass that demonstrated in other studies for small animal imaging, it presents a potential solution for whole-body human imaging of the TAT agent. It is important to note that the current evaluation only considers two plate detectors; ongoing improvements in system geometry will be conducted. Further evaluation of imaging performance will involve moving the detector along the patient bed while simulating a body phantom, which necessitates an accurate design of the patient body phantom.

## Acknowledgements

All authors declare that they have no known conflicts of interest in terms of competing financial interests or personal relationships that could have an influence or are relevant to the work reported in this paper.

## Appendix I

List of Symbols

L: Distance from source to detector's scatterer surface
d: Distance between scatterer and absorber
$b_1$, $b_2$: Thickness of scatterer and absorber
$dL_1$, $dL_2$: Size of side of scatterer and absorber
$L_1$, $L_2$: Distance from source to first and second detector head
k: Material density
M: Center of the voxel
$V_1$, $V_2$: Detected locations
$E_0$, $E_1$, $E_2$: Initial energy and deposited energies
$\beta_G$: Angle between vector $\overrightarrow{V_1M}$ and cone's central axis
$\beta_E$: Compton scattering angle determined by energies
$\rho$: Conical uncertainties in cm
$\Delta\beta$: Angle uncertainties
$\delta$: Angle difference between $\beta_E$, $\beta_G$.
$p_1$, $p_2$: Pixel size of scatterer and absorber
$V_j$: Volume of voxel j
i: One event in the Compton event list mode data
j: Voxel j
$\lambda_j$: Reconstructed emission at voxel j
$S_j$: Probability of photon emitted from voxel j being detected
$t_{ij}$: Probability of detected event i being emitted form voxel j
T: System matrix
S: Sensitivity matrix
$\lambda$: Reconstructed image
$\mu_c$: Linear attenuation coefficient for Compton
$\theta_j$: Angle between the voxel and detector pixel